\documentclass[10pt,conference]{IEEEtran}
\usepackage{cite}
\usepackage{amsmath,amssymb,amsfonts}
\usepackage{algorithmic}
\usepackage{graphicx}
\usepackage{textcomp}
\usepackage{xcolor}
\usepackage[hyphens]{url}
\usepackage{fancyhdr}
\usepackage{hyperref}

\newcommand{\hpcayear}{2026}

\usepackage{amsmath}
\usepackage{verbatim} % for \begin{comment}
\usepackage{stfloats} % allow figure* at bottom of page
\usepackage{subcaption} % for subfigure
\usepackage{epsfig} % to set graphics path
\usepackage{xspace}
\graphicspath{{./figures/}}

\usepackage{makecell}
\usepackage{booktabs}
\usepackage{array}

\usepackage[skip=4pt]{caption} % make figures dense
\usepackage{caption}
\DeclareCaptionFont{ninept}{\fontsize{9pt}{10pt}\selectfont}
\usepackage{microtype} % margin between letters
\newcolumntype{L}[1]{>{\raggedright\arraybackslash\vspace{0pt}}p{#1}}
\newcolumntype{C}[1]{>{\centering\arraybackslash\vspace{0pt}}p{#1}}
\newcolumntype{R}[1]{>{\raggedleft\arraybackslash}p{#1}}

\usepackage{url}
\usepackage{hyperref} % should be last package
\usepackage{cleveref} % should be after hyperref
\crefname{figure}{Figure}{Figures}
\crefname{table}{Table}{Tables}
\crefname{section}{Section}{Sections}
\crefname{chapter}{Chapter}{Chapters}
\crefname{equation}{Equation}{Equations}

\usepackage{listings}
\usepackage{multicol} % in preamble

\newcommand{\smalltt}[1]{{\fontsize{9}{11}\selectfont\texttt{#1}}}

\newcommand{\ada}[1]{\textcolor{red}{AG:\ #1}}

\newenvironment{tightitemize}%
{
  \begin{list}{$\bullet$}{%
      \setlength{\leftmargin}{10pt}
      \setlength{\itemsep}{0pt}%
      \setlength{\parsep}{0pt}%
      \setlength{\topsep}{0pt}%
      \setlength{\parskip}{0pt}%
    }%
    }%
    {
  \end{list}
}
\newcounter{tecounter}
\newenvironment{tightenumerate}%
{
  \begin{list}{\arabic{tecounter}.}{%
      \usecounter{tecounter}
      \setlength{\leftmargin}{10pt}
      \setlength{\itemsep}{0pt}%
      \setlength{\parsep}{0pt}%
      \setlength{\topsep}{0pt}%
      \setlength{\parskip}{0pt}%
    }%
    }%
    {
  \end{list}
}%

{
  \begin{list}{}{%
      \setlength{\leftmargin}{10pt}
      \setlength{\itemsep}{0pt}%
      \setlength{\parsep}{0pt}%
      \setlength{\topsep}{5pt}%
      \setlength{\parskip}{0pt}%
    }%
    }%
    {
  \end{list}
}%

\newcommand{\sys}{\mbox{\textsc{Rocket}}\xspace}

\title{Rethinking Inter-Process Communication with Memory Operation Offloading}
\newcommand\hpcaauthors{First Author$\dagger$ and Second Author$\ddagger$}
\newcommand\hpcaaffiliation{First Affiliation$\dagger$, Second Affiliation$\ddagger$}
\newcommand\hpcaemail{Email(s)}

\def\hpcaarxiv{}           % Uncomment to build arXiv/archive version
\newcommand{\sethpcaauthors}{%
  \renewcommand\hpcaauthors{%
    Misun Park\textsuperscript{$\dagger$}, Richi Dubey\textsuperscript{$\dagger$}, Yifan Yuan\textsuperscript{$\ddagger$}, Nam Sung Kim\textsuperscript{$\S$}, Ada Gavrilovska\textsuperscript{$\dagger$}%
  }%
  \renewcommand\hpcaaffiliation{%
    Georgia Institute of Technology\textsuperscript{$\dagger$}, Meta\textsuperscript{$\ddagger$}, University of Illinois--Urbana-Champaign\textsuperscript{$\S$}%
  }%
  \renewcommand\hpcaemail{%
  \texttt{misun@gatech.edu}, 
  \texttt{richidubey@gatech.edu}, 
  \texttt{yifanyuan@meta.com}, \\
  \texttt{nskim@illinois.edu}, 
  \texttt{ada@cc.gatech.edu}%
}
}

\ifdefined\hpcaarxiv
  \sethpcaauthors
\fi

\author{
  \ifdefined\hpcacameraready
    \IEEEauthorblockN{\hpcaauthors{}}
      \IEEEauthorblockA{
        \hpcaaffiliation{} \\
        \hpcaemail{}
      }
  \else\ifdefined\hpcaarxiv
    \IEEEauthorblockN{\hpcaauthors{}}
    \IEEEauthorblockA{
      \hpcaaffiliation{} \\
      \hpcaemail{}
    }
  \else
    \IEEEauthorblockN{\normalsize{HPCA \hpcayear{} Submission
      \textbf{\#\hpcasubmissionnumber{}}} \\
      \IEEEauthorblockA{
        Confidential Draft \\
        Do NOT Distribute!!
      }
    }
  \fi\fi 
}

\fancypagestyle{camerareadyfirstpage}{%
  \fancyhead{}
  
  \fancyhead[C]{
    \ifdefined\aeopen
    \parbox[][12mm][t]{13.5cm}{\hpcayear{} IEEE International Symposium on High-Performance Computer Architecture (HPCA)}    
    \else
      \ifdefined\aereviewed
      \parbox[][12mm][t]{13.5cm}{\hpcayear{} IEEE International Symposium on High-Performance Computer Architecture (HPCA)}
      \else
      \ifdefined\aereproduced
      \parbox[][12mm][t]{13.5cm}{\hpcayear{} IEEE International Symposium on High-Performance Computer Architecture (HPCA)}
      \else
      \parbox[][0mm][t]{13.5cm}{\hpcayear{} IEEE International Symposium on High-Performance Computer Architecture (HPCA)}
    \fi 
    \fi 
    \fi 
    \ifdefined\aeopen 
      \includegraphics[width=12mm,height=12mm]{ae-badges/open-research-objects.pdf}
    \fi 
    \ifdefined\aereviewed
      \includegraphics[width=12mm,height=12mm]{ae-badges/research-objects-reviewed.pdf}
    \fi 
    \ifdefined\aereproduced
      \includegraphics[width=12mm,height=12mm]{ae-badges/results-reproduced.pdf}
    \fi
  }
  \fancyfoot[C]{}
}
\begin{document}
\maketitle

%Enables the camera ready header and footer
\ifdefined\hpcacameraready 
  \thispagestyle{camerareadyfirstpage}
  \pagestyle{empty}
\else
  \thispagestyle{plain}
  \pagestyle{plain}
\fi

\newcommand{\hpcaheight}{0mm}
\ifdefined\eaopen
\renewcommand{\hpcaheight}{12mm}
\fi

%%%%%%%%%%%%%%%%%%%%%%%%%%%%%%%%%%%%%%%%
%%%%%%%% -- PAPER CONTENT STARTS -- %%%%%%%%%

% We suggest an architecture for a system~\cite{satyanarayanan1990coda}.
\begin{abstract}

As multimodal and AI-driven services exchange hundreds of megabytes per request, existing IPC runtimes spend a growing share of CPU cycles on memory copies. Although both hardware and software mechanisms are investigating memory offloading, current IPC stacks lack a unified runtime model to coordinate them effectively.
%\newline
This paper presents \sys, an IPC runtime suite that integrates both hardware- and software-based memory offloading into shared-memory communication. \sys systematically characterizes the interaction between offload strategy %engine  
and IPC execution, covering synchronization, cache visibility, and concurrency, and introduces %adaptive 
different IPC
modes that balance throughput, latency, and CPU efficiency. 
%\newline
Through asynchronous pipelining, selective cache injection, and hybrid coordination, \sys turns offloading from a device-specific feature into a general system capability.
Evaluations on real-world workloads show that \sys reduces instruction counts by up to 22\%, improves throughput by up to 2.1$\times$, and lowers latency by up to 72\%, demonstrating that coordinated IPC %memory 
offloading can deliver tangible end-to-end efficiency gains in modern data-intensive systems.

\end{abstract}

\section{Introduction}\label{sec:intro}

  %<sigchi-a>\end{margintable}

Despite massive progress in compute acceleration, modern AI and data analytics pipelines are increasingly bottlenecked not by computation, but by inter-process data movement. Emerging multimodal workloads, such as video, image-text, or tabular, routinely exchange hundreds of megabytes per request~\cite{lotus,klimovic:plumber,klimovic:socc23}. For instance, 
batching 256 RGB images (224×224, fp32) already moves over a hundred MB per offline inference, and multimodal or visual analytics pipelines further amplify these transfers.
A single 4K image exceeds 30MB~\cite{mri-dicom}, and multimodal pipelines push request sizes even higher.
These figures correspond to input sizes frequently seen in AI inference pipelines, large-scale graph queries, and tabular data processing workloads~\cite{minibatch-256,cvpr-big-batch,4kimage,hadoop}.
\autoref{tab:data-movement} summarizes this trend by showing the characteristics of several representative multimodal workloads. These workloads now involve inter-process transfers of 50-500 MB, turning memory copy into the dominant latency and energy cost.

\begin{table}[t]
\centering
\caption{Metrics characterizing data transfer and memory behavior for representative multimodal workloads in application-pipeline IPC scenarios.}
\label{tab:data-movement}
\footnotesize
\setlength{\tabcolsep}{2pt}
\begin{tabular}{
    p{1.7cm}  % Aspect
    >{\arraybackslash}p{1.55cm}  % MobileNetV2
    >{\arraybackslash}p{1.55cm}  % XGBoost
    >{\arraybackslash}p{1.75cm}  % PageRank
    >{\arraybackslash}p{1.35cm}  % MilvusDB
}
\toprule
\textbf{Aspect} & \textbf{MobileNetV2} & \textbf{XGBoost} & \textbf{PageRank} & \textbf{MilvusDB} \\
\midrule
\textbf{Bytes\newline(req/resp)} & 120MB/\newline800KB & 25MB/\newline800KB & 76MB/\newline320KB & 1MB/\newline320MB \\
\midrule
\textbf{Memcpy time\newline in IPC (ms)}  & 203 & 25 & 46 & 323 \\
\midrule
\textbf{Memory\newline behavior} &
Large input,\newline low reuse &
Low reuse,\newline compute-bound &
High reuse,\newline memory-bound &
Bulky\newline response \\
\midrule
\textbf{Config} &
Batch size\newline200~\cite{shvit} &
200K rows,\newline30 features &
Graph w/\newline10M edges~\cite{snapnets} &
Batched\newline search \\
\bottomrule
\end{tabular}
\vspace{-5.0ex}
\end{table}

% In iterative stages in AI training or inference pipelines, data exchange, not computation, has become the limiting factor, saturating memory bandwidth and CPU cycles even before the core task begins (\cite{mlsys2021}).

% While multi-node systems process massive datasets, even intra-node pipelines must efficiently move data across components like %databases 
% data caches~\cite{cachew}
% and inference backends which share accelerators~\cite{pocket, orion:klimovic}. 

% In serverless workflows, up to 92\% of end‑to‑end latency stems from data movement within a node~\cite{faastube}, and scientific scatter-gather pipelines routinely exchange 32MB to 512MB chunks within a node while scaling experiments up to 1GB transfers~\cite{dhmem}.
% The growing size and frequency of data batches in modern pipelines increase intra-node memory transfers, a trend intensified by heterogeneous nodes integrating GPUs, FPGAs, and ASICs, resulting in higher data traffic and contention across memory and I/O subsystems~\cite{intranode}.

Prior research 
reported that even with pre-AI datacenter workloads, in-node data transfer (\smalltt{memmove}) consumes over 5\% of total CPU cycles, contributing to what is often referred to as the data center tax~\cite{isca-datacenter},
incurring secondary costs such as cache pollution and energy overheads~\cite{isca-datacenter}.
With the popularity of multimodal AI, data-intensive workloads continue to expand~\cite{realtime-ml}, making efficient inter-process communication (IPC) mechanisms and memory copy within a data center node play a critical role in achieving scalable and high-performance system architectures.
\begin{figure}[t]
  \centering
  \includegraphics[width=0.9\columnwidth]{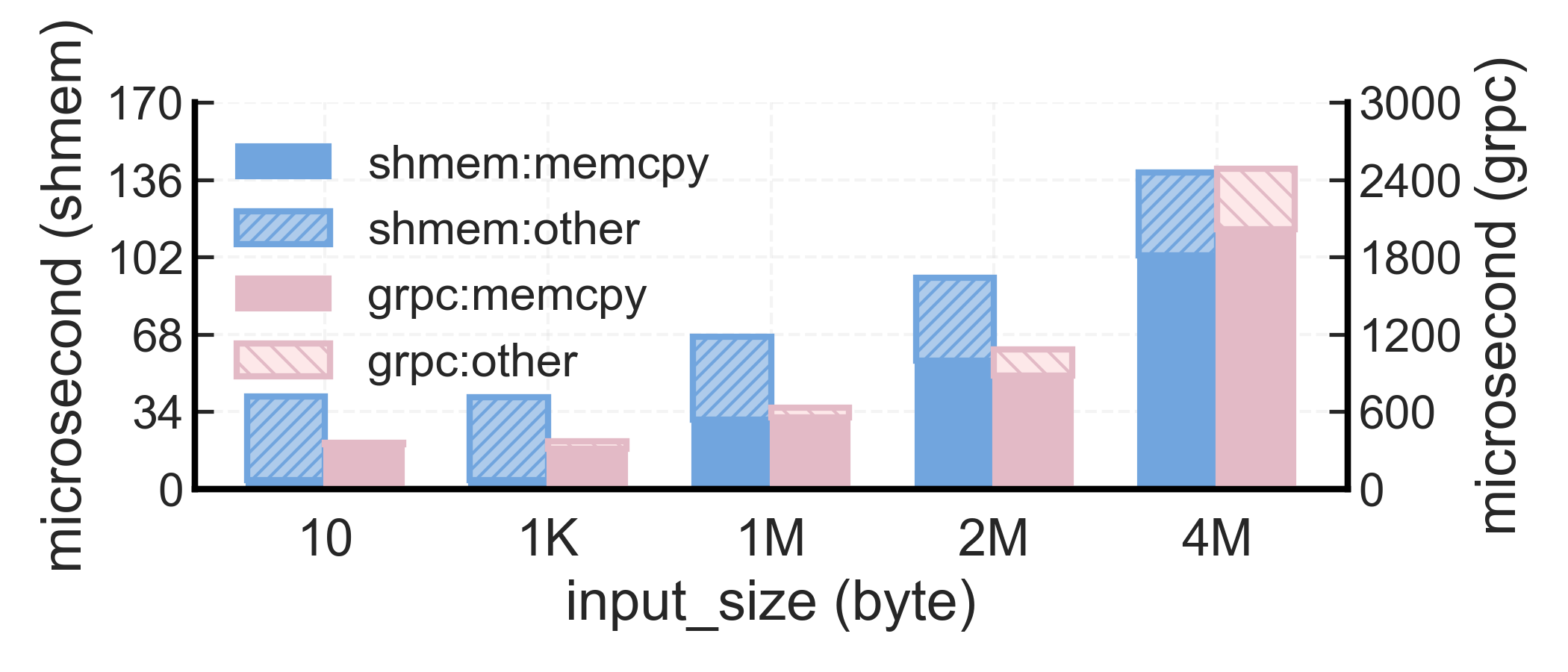}
  % \includegraphics[width=0.9\columnwidth]{figures/memopdist-nop.png}
  % \vspace{-1.0ex}
  \caption{Breakdown of end-to-end latency for intra-node echo RPCs implemented using shared memory (shmem) and gRPC. The figure quantifies the portion of total latency attributed to \smalltt{memcpy} as a function of message size.}
  \vspace{-4.0ex}
  \label{fig:memcpyanatomy}
\end{figure}
\autoref{fig:memcpyanatomy} shows that  \smalltt{memcpy} time dominates the execution of both optimized (i.e., shared-memory-based implementations such as Nightcore~\cite{nightcore}) and general-purpose IPC stacks (e.g., gRPC), even at moderate data sizes. 
As data transfer sizes increase, curtailing these memory data transfer costs can have major impact on data center efficiency. 

Existing IPC stacks, designed around CPU-driven memcpy, cannot sustain such transfer volumes without saturating cores and cache hierarchies. This growing gap between compute and communication efficiency motivates the need for a new class of offload-aware IPC runtimes. 
Emerging hardware engines for data acceleration~\cite{idma} expose an opportunity to offload these transfers from CPUs, freeing up CPU resources for application tasks and reducing stalls from memory overhead. Recent system-level work~\cite{copier} demonstrates that even software-based offload mechanisms integrated as OS services can  improve copy performance, highlighting the increasing importance of offload-aware design across hardware and software layers.

Prior work on memory offloading has primarily examined \textit{intra-process} data movement using microbenchmarks and database workloads~\cite{dimes-dsa,asplos-kuper,damon-25}, as well as for network-stack acceleration in \textit{inter-node} settings~\cite{asplos-kuper}. Yet, integrating such offload into IPC stacks introduces subtle trade-offs due to cache interference and synchronization overhead that existing systems cannot handle. This is critical in inter-process, intra-node settings -- central to client-server interactions~\cite{deathstar} and emerging data processing pipelines~\cite{lotus,klimovic:plumber,klimovic:socc23} --  yet, remains underexplored. 

Integrating memory accelerators into inter-process data movement introduces new performance pitfalls. Most IPC systems still rely on synchronous communication semantics, which restrict the parallelism that hardware offload can provide. Without designs aware of such accelerators, naïve integration often breaks cache locality, disturbs memory access patterns, and amplifies page fault and translation overheads. Instead of accelerating execution, these effects can negate the benefits of offload and even degrade overall performance.

This work investigates how data-movement accelerators integrate into shared-memory IPC pipelines for multi-process applications. Intel’s DSA serves as a concrete observation point to study this integration in realistic, memory-intensive workloads. We characterize the system-level factors that shape accelerator efficiency in state-of-the-art IPC runtimes, identifying when offload improves performance and when it introduces new bottlenecks. Our analysis reveals key dimensions, such as cache behavior, synchronization cost, contention, page faults, and execution model, that determine how well hardware offload aligns with software.

Building on these findings, we design \sys, an IPC software stack that optimizes accelerator interaction along these dimensions. The results show that performance is highly sensitive to integration choices: \sys reduces instruction count by up to 22\% and increases throughput by 15\%.

The contributions of this work can be summarized as follows:

\noindent
\textbf{(1) System-Level Bottleneck Analysis.}\hspace*{0.5em}
We identify key performance factors affecting hardware-based memory offloading in user-space IPC, including CPU usage, cache behavior, and synchronization overhead. This analysis reveals fundamental bottlenecks that limit accelerator effectiveness in real deployments. (\S\ref{sec:tradeoffs})

\noindent
\textbf{(2) Protocol Design and Implementation.}\hspace*{0.5em}
We design and implement a suite of shared-memory IPC protocols -- \sys -- that integrate memory offloading into user-space pipelines. The design reflects key integration dimensions such as synchronization models, execution modes, and cache behavior. \sys improves efficiency through asynchronous pipelining, parallelism, and cache injection, while utilizing power-saving instructions on x86. (\S\ref{sec:design})

\noindent
\textbf{(3) Empirical Evaluation.}\hspace*{0.5em}
We evaluate \sys on real workloads on a system with an Intel DSA offload engine, and show up to 22\% fewer instructions and 15\% higher throughput compared to a CPU baseline, outperforming current software support. These gains extend beyond memcpy speedups, enabled by offload-aware execution paths and integration-conscious pipeline structuring. (\S\ref{sec:eval})

\noindent
This work provides both a practical IPC design for memory engines and a systematic understanding of the architectural trade-offs in user-space deployment, laying the groundwork for future memory-accelerated systems in data-intensive environments.

\section{Background: Hardware-Assisted Memory Offloading}\label{sec:background}
% \misun{I'm trying to reframe rocket more general, . previous version sounded like DSA tutorial, esp this bg section.}

Modern systems increasingly seek to decouple data movement from CPU execution. Conventional memcpy-based transfers, though simple and universal, become costly under bandwidth-intensive workloads because each copy pollutes caches and consumes valuable CPU cycles. To mitigate these inefficiencies, hardware and operating systems now treat memory offload as a programmable service, allowing bulk transfers to proceed outside the CPU's critical path. Recent efforts, from kernel-managed asynchronous copy engines to integrated accelerators such as Intel's DSA, reflect this shift and motivate a deeper exploration of software-hardware integration in runtime systems.

\subsection{Landscape in Memory Offloading}
In recent years, memory offloading has become a prominent trend in system design. Copy operations, once treated as simple CPU-bound library calls, now emerge as key performance bottlenecks under data-intensive workloads. Cache interference, synchronization delays, and wasted CPU cycles limit scalability. To address these issues, modern systems increasingly delegate copy operations to specialized hardware or kernel-level services.

At the hardware level, recent work performs copies at the memory controller or DMA engine \cite{mc2,dmx}, reducing CPU stalls and enabling transfers to overlap with computation. At the kernel and system level, copy operations are reinterpreted as coordinated services 
and delegated to designated threads
%rather than incidental instruction effects 
\cite{copier}. These systems track dependencies between copy requests and overlap copy-use phases to improve throughput. Collectively, they aim to make data movement a first-class, schedulable operation.
Yet these efforts largely focus on \emph{enabling} memory offload, not on
\emph{integrating} it into application-level dataflows.  
The resulting gap, between what hardware or kernel mechanisms can do and what
end-to-end systems actually exploit, defines where Rocket operates.  
Our goal is to articulate this missing software layer: a runtime that bridges
offload-capable hardware and user-space IPC.

%\ada{is the point you're trying to make here that it's necessary to carefully consider the integration of these mechanisms (regardless of level of implementation) in a use case-specific manner?} 
%\misun{the point that i'm trying to make here is, rocket work is in this big current, but even in this current there's some missing piece, application level integration, and we're doing it. I re wrote the last sentence. can you read it?}

\autoref{tab:offload-comparison} summarizes these developments and their trade-offs across hardware, kernel, and system layers, highlighting how cache management, synchronization, and policy flexibility remain key design constraints.

\begin{table}[h!]
  \centering
  \small
  \caption{Representative memory-offload mechanisms across system layers.} % \ada{is it DMC or DMX?}}
  \label{tab:offload-comparison}
  \begin{tabular}{
    p{1.6cm}  % Category
    p{1.8cm}  % MC2
    p{1.8cm}  % DMX
    p{1.8cm}  % Copier
  }
    \toprule
    & \textbf{(MC)$^2$~\cite{mc2}} & \textbf{DMX~\cite{dmx}} & \textbf{Copier~\cite{copier}} \\
    \midrule
    \textbf{Level} &
      Mem. controller &
      Cross-accelerator &
      OS kernel \\
    \midrule
    \textbf{Cache\newline pollution} &
      Lazy flush &
      Fence DMA &
      Not handled \\
    \midrule
    \textbf{Applicability} &
      Hardware-tied &
      Specialized setup &
      OS-integrated \\
    \midrule
    \textbf{Overhead} &
      Very low &
      Moderate &
      Moderate-high \\
    \bottomrule
  \end{tabular}
\end{table}

\subsection{Case Study: Intel DSA} 
A representative example of modern memory offloading engines is Intel's DSA, integrated into Sapphire Rapids processors. DSA extends traditional DMA with user-level control, virtual memory support, and fine-grained cache management. These features aim to reduce cache pollution, free CPU cycles, enable compute-memory overlap, and improve bandwidth utilization.

%\noindent
%\textbf{vs. Traditional DMA.}\hspace*{0.5em} DSA introduces several key improvements over traditional DMA engines that make it a more powerful and versatile solution for handling memory-bound workloads.

\noindent
\textbf{Virtual Memory Support.}\hspace*{0.5em}
Unlike traditional DMA that require physical addresses, DSA supports virtual memory, simplifying integration with 
% reducing the complexity of integrating DSA into 
user-space applications.
% This capability is particularly important in virtualized environments and microservices, where physical memory-based DMA is difficult to apply. DSA overcomes this limitation through
% DSA relies on
It leverages PASID (Process Address Space ID) to manage multiple address spaces without explicit memory pinning, making it well-suited for virtualized and multitenant environments.
% support to enable seamless management of multiple address spaces without requiring explicit memory pinning. %This makes DSA suitable for cloud environments with virtualized and multi-tenant workloads, where efficient memory movement and address space isolation are critical.

\noindent
\textbf{Enhanced Programmability.}\hspace*{0.5em}
A key advantage of DSA is its improved programmability. Unlike traditional DMA engines that rely on system calls (e.g., \smalltt{ioctl}) and kernel-managed descriptors, DSA allows direct submission of work descriptors from user space via dedicated CPU instructions such as \smalltt{ENQCMD} and \smalltt{MOVDIR64B}. This avoids context switches and kernel entry, reducing latency from tens of microseconds~\cite{syscall-overheads} to a few hundred CPU cycles ($\approx$ 200ns)\cite{dsa-userguide}.
Additionally, \smalltt{ENQCMD} executes atomically\cite{asplos-kuper}, eliminating the need for locks in multithreaded contexts.
% Another advantage is that configuration parameters -- such as queue count, queue size, and cache injection policy -- can be adjusted entirely from user-space via the \smalltt{accel-config} interface, without needing kernel-level reconfiguration.

% \noindent
% \textbf{Reduced Need for Memory Pinning.}\hspace*{0.5em} By supporting non contiguous memory and virtual memory addressing, DSA reduces the need for memory pining, which has been a significant constraint for DMA engines. This allows for more efficient use of system memory, especially in workloads that involve large datasets. % or frequent memory access.
\vspace{-1.0ex}
\subsection{Software Stack for DSA}

We examine two approaches to using DSA: the low-level interface provided by the DSA driver and Intel’s DSA Transparent Offload (DTO) framework. The former offers fine-grained control and low-latency paths for advanced users, while the latter prioritizes ease of use via library call interception, trading off flexibility and performance. These implementations illustrate the trade-offs between programmability, control, and performance in current DSA software support.

\begin{figure}[t]
    \centering
    \includegraphics[width=0.90\columnwidth]{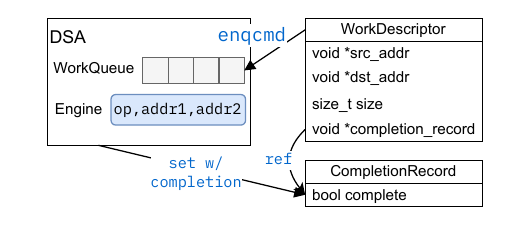}
    \caption{DSA programming model. The CPU prepares the task descriptor and submits it to DSA. DSA executes the task and sets the completion flag. The CPU then checks the completion flag to determine if the task is complete.}
    \vspace{-3.0ex}
    \label{fig:how-to-use-dsa}
\end{figure}

\noindent
\textbf{Low-level Programming Interface.}\hspace*{0.5em}
The DSA programming model involves three main steps. First, the CPU prepares a task descriptor that specifies the memory operation to offload, including source and destination addresses, transfer size, and a completion flag address. Second, the descriptor is submitted to the DSA device via low-level enqueue instructions. This step requires direct interaction with hardware-specific structures exposed by \smalltt{libaccel-config.h} and \smalltt{linux/idxd.h}~\cite{dsa-userguide}.
Third, upon completion, the DSA sets the flag, which the CPU can monitor either by polling or asynchronously via mechanisms like \smalltt{UMWAIT}. While interrupt-based completion is considered to be the most efficient methods, it is not available in user mode due to the lack of interrupt handling capabilities.
This offload model reduces CPU involvement during data movement, freeing cycles for other tasks (\autoref{fig:how-to-use-dsa}).

%Unlike traditional DMA programming, submitting a task to the DSA does not require a system call such as \smalltt{ioctl()}. Instead, the descriptor is written directly to a memory-mapped work queue using the \smalltt{ENQCMD} instruction, allowing efficient and low-latency submission from user space. This simplifies the programming model and reduces kernel involvement, which is especially advantageous in latency-sensitive applications.

%Although interrupt-based completion checking is generally regarded as the most efficient method, it is not feasible in user-mode applications due to the lack of interrupt handling capabilities. As a result, polling the completion flag remains necessary, introducing potential inefficiencies depending on how and when the polling is performed.

Other interfaces, such as the Storage Performance Development Kit (SPDK), provide user-level libraries for working with DSA. However, these are often thin wrappers around the core programming model described above and do not fundamentally alter the task submission or completion semantics.

\noindent
\textbf{Static Offloading.}\hspace*{0.5em}
Intel's DSA Transparent Offload (DTO) framework allows applications to use DSA without source code changes by intercepting standard library calls such as \smalltt{memcpy()} and redirecting them to DSA. This simplifies adoption, especially for legacy or closed-source software.

However, DTO has two key limitations.
(1) It lacks fine-grained control over offload decisions: all intercepted calls are offloaded uniformly, regardless of transfer size, locality, or reuse distance. This can hurt performance when CPU-based \smalltt{memcpy()} is preferable (e.g., for small, latency-sensitive transfers).
(2) DTO enforces a synchronous execution model, preventing pipelining or overlap between DSA transfers and CPU computation, limiting parallelism.

These constraints make DTO easy to adopt but less suitable for performance-critical scenarios, motivating alternative programming models.

\section{Motivation: Integration Challenges in Offloaded IPC}
\label{sec:tradeoffs}

% \ada{Overall this is section has a good flow. There is some repetition, both within this section and also with later in design again some points are repeated or not coming across. Table should be revisited, along the lines discussed in last meeting, maybe adding a column on what's the implication on design requirement for \sys.}

% \ada{The key thing that's missing is some data points to better substantiate the statements about the tradeoffs. Can you include some microbenchmark results? You probably have some data in the results you already have. e.g., around the fact that cache infection is sometimes good (speeds things up, reduces cache misses, but sometimes bad, has the oposite effect, but to what extent); around polling/synchronization issue related to cache misses and stalls vs. use of user mode interrupts; also that you want to control whether you use DSA sync or stay on CPU, like if you use with small data vs. large data (you have this data I believe). }

Incorporating
% \ada{remove hardware-assisted? offload/delegation to different threads in copier may raise similar tradeoffs (at a very different scale...)} hardware-assisted 
memory offloading into inter-process communication introduces subtle trade-offs that can negate its benefits when applied naively. Offload engines relieve the CPU of data movement but shift synchronization, visibility, and caching responsibilities to software. This interplay exposes three recurring tensions: (i) synchronization granularity between submission and completion, (ii) address visibility and page-fault handling, and (iii) cache injection and data reuse control. Understanding these tensions is essential for integrating offload mechanisms into runtime systems effectively and motivate the design principles in \S\ref{sec:design}. 
In this section, we illustrate and quantify these tradeoffs for Intel DSA.

% While DSA offers significant advantages over traditional DMA engines, effectively implementing memory operation offloading is challenging. A naive implementation can lead to unintended side effects, such as disrupting the existing memory hierarchy and causing performance degradation. In such cases, the benefits of DSA may be significantly limited, falling short of theoretical maximum improvements, or even outweighed by the inefficiencies it introduces.

% \begin{figure}[h!]
%   \centering
%   \includegraphics[width=0.9\columnwidth]{figures/dto-ipc.png}
%   \caption{IPC performance comparison between DSA offloaded memcpy and CPU-based memcpy. The actual performance improvement is much more limited than the theoretical expectation.}\label{fig:dto-ipc}
% \end{figure}
% (\autoref{fig:dto-ipc})

% In this section, we first analyze and present the hardware-level benefits and costs associated with the introduction of DSA. Second, we examine how memory copy operations, which are conventionally executed by the CPU, follow a different execution path when performed through DSA. This deviation challenges our traditional assumptions regarding the memory hierarchy and the philosophy of temporal locality, leading to significant implications. Finally, based on these insights, we derive the design requirements necessary for developing optimal software support for DSA.

\subsection{Hardware-level Trade-offs}

Offloading memory operations frees CPU cycles and reduces cache pollution, potentially improving overall performance. For example, offloading a 1MB transfer saves about 33$\mu s$ with DSA($\approx$130,000 CPU cycles that can be repurposed for other tasks). However, these benefits are context-dependent and may incur system-level overhead if not carefully managed.

%
%\ada{this was said in previous section, remove}
%Intel introduced the Data Streaming Accelerator (DSA) as an advanced alternative to traditional Direct Memory Access (DMA), first integrating it into their Sapphire Rapids processors. While DMA has long been used to offload memory operations from the CPU, DSA enhances this approach by offering superior programmability, flexibility, and overall functionality. The key benefits of DSA include reduced cache pollution, the ability to free up CPU cycles, improved performance through parallel execution of computation and memory operations, and reduced overall memory bandwidth pressure.
In IPC workloads, DSA integration presents several hardware-level trade-offs. Key factors include synchronization overhead from busy-waiting, performance loss from bypassing CPU caches, latency spikes from page faults, and potential bus contention. The following sections examine each and its impact on system behavior.

% Compared to traditional DMA, DSA provides several key improvements:

% \begin{tightitemize}
%     \item \textbf{Virtual Memory Support}\hspace*{0.5em} Unlike DMA, which operates solely on physical addresses, DSA supports virtual memory addresses, making it easier to integrate into user-space applications.
%     \item \textbf{Enhanced Programmability}\hspace*{0.5em} DSA offers a more flexible API, allowing developers to issue commands via CPU instructions rather than manually configuring registers, thereby improving ease of use.
%     \item \textbf{Reduced Need for Memory Pinning}\hspace*{0.5em} DSA's ability to work with non-contiguous memory and virtual memory addressing minimizes the need for memory pinning, which has been a major constraint in traditional DMA implementations.
%     % This enables more efficient memory management, particularly for workloads involving large datasets and frequent memory access.
% \end{tightitemize}

% Although DSA can sequentially execute memory operations on behalf of the CPU, allowing the CPU to perform other tasks in parallel, simply offloading memory operations does not automatically guarantee performance gains. While DSA-driven memcpy operations are generally faster than their CPU-based counterparts, offloading to an external device introduces the additional complexity of tracking command completion, resulting in synchronization overhead.

\begin{figure}[t]
  \centering
  \hspace{-2.5ex}
  \includegraphics[width=0.90\columnwidth]{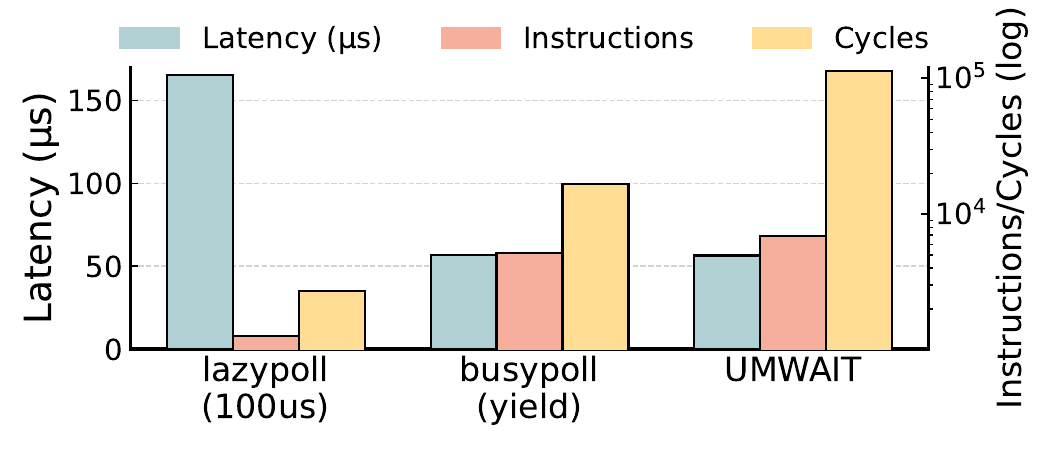}
  \caption{Comparison of polling strategies on latency and CPU usage (1MB transfer. \smalltt{lazypoll}: polling every 100$\mu$s; \smalltt{busypoll}: polling with \texttt{yield} but no sleep; \smalltt{UMWAIT}: polling with usermode interrupt.)}
  \vspace{-3.0ex}
  \label{fig:polling-strategy}
\end{figure}

% \vspace{\baselineskip}
% \misun{does this look good enough?}
\noindent
\textbf{Overheads from Completion Check.}\hspace*{0.5em}
% As illustrated in \autoref{fig:how-to-use-dsa}, the DSA programming model consists of three steps: (1) the CPU prepares a task descriptor specifying the memory operation and a completion flag address, (2) the descriptor is submitted to the DSA engine, and (3) upon completion, DSA writes to the flag address.
Regardless of whether DSA is used synchronously or asynchronously, offload completion must be detected by reading a completion flag in an uncacheable memory-mapped I/O region.

\autoref{fig:polling-strategy} compares three polling strategies. Busy-waiting provides low latency but consumes high CPU cycles. Lazy-waiting is inefficient in latency. \smalltt{UMWAIT} offers latency comparable to busy-waiting, but does not provide true sleep or asynchronous behavior—it places the CPU in a shallow wait state, effectively polling at 25µs intervals. In single-threaded settings, its main benefit is power savings rather than responsiveness\cite{intel-dpdk}.

Polling introduces nontrivial system-level costs. Each read to the uncacheable flag bypasses the CPU cache and traverses the memory bus, increasing contention. Additionally, accesses to memory-mapped I/O regions enforce strict ordering constraints, hindering out-of-order execution and introducing pipeline stalls. These accesses may also cause cache and TLB invalidations, degrading overall performance~\cite{polling-mem-bandwidth, pipeline-stalls}.
While memory offloading reduces CPU involvement in data movement, frequent polling for synchronization can offset its benefits. This highlights the need for low-overhead, responsive synchronization mechanisms.
Using DSA in synchronous mode increases synchronization overhead and extends CPU idle time. Asynchronous mode can reduce this cost, making it preferable for efficient offload. Still, due to the nature of accelerator execution, completion checking cannot be entirely avoided, and minimizing its overhead remains a key challenge.

\begin{figure}[t]
   \centering
   % \hspace{-2.5em}
   \includegraphics[width=0.95\columnwidth]{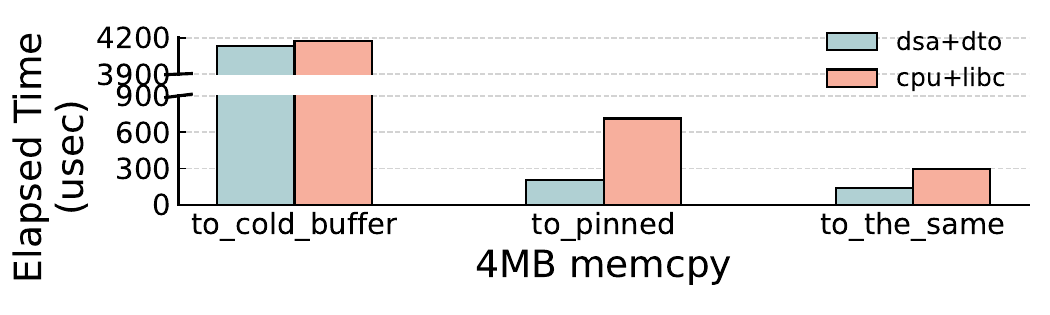}
   \vspace{-1.0ex}
   \caption{Performance comparison of DSA and CPU \smalltt{memcpy} under different memory conditions. 
Copying to a pinned buffer reduces latency by 95\%, and reusing the same buffer achieves a 97\% reduction, both relative to cold-buffer access.}
    \vspace{-3.0ex}
   \label{fig:pagefaults}
\end{figure}

\begin{figure}[t]
  \centering
  \includegraphics[width=0.85\columnwidth]{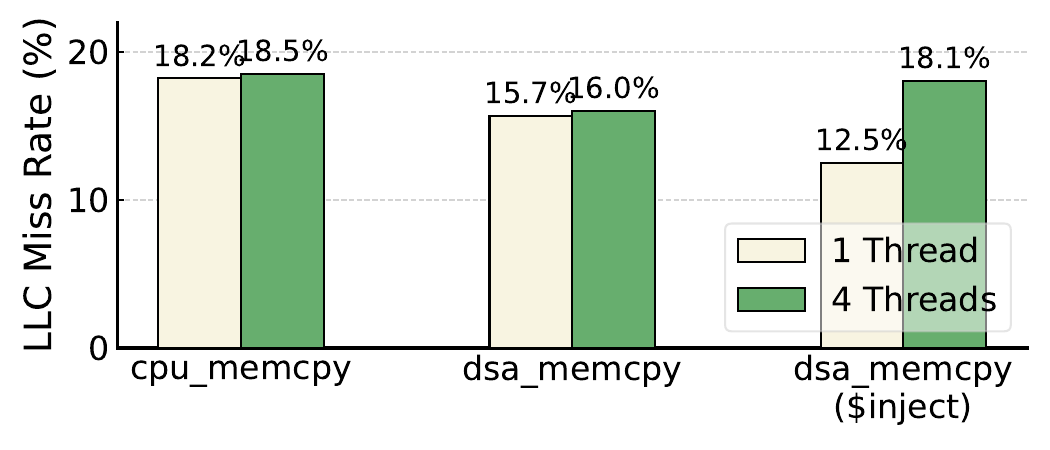}
  \vspace{-1.5ex}
  \caption{LLC miss rates under single and four-threaded execution, comparing \smalltt{cpu\_memcpy}, \smalltt{dsa\_memcpy}, \smalltt{dsa\_memcpy(\$inject)} (Microbenchmark: summation over all elements in the destination buffer after memory copy).}
  \vspace{-4.0ex}
  \label{fig:cache-injection}
\end{figure}

\begin{figure*}[b]
  \vspace{-2.0ex}
  \centering
  \begin{subfigure}[b]{0.4\linewidth}
      \centering
      \includegraphics[width=\linewidth]{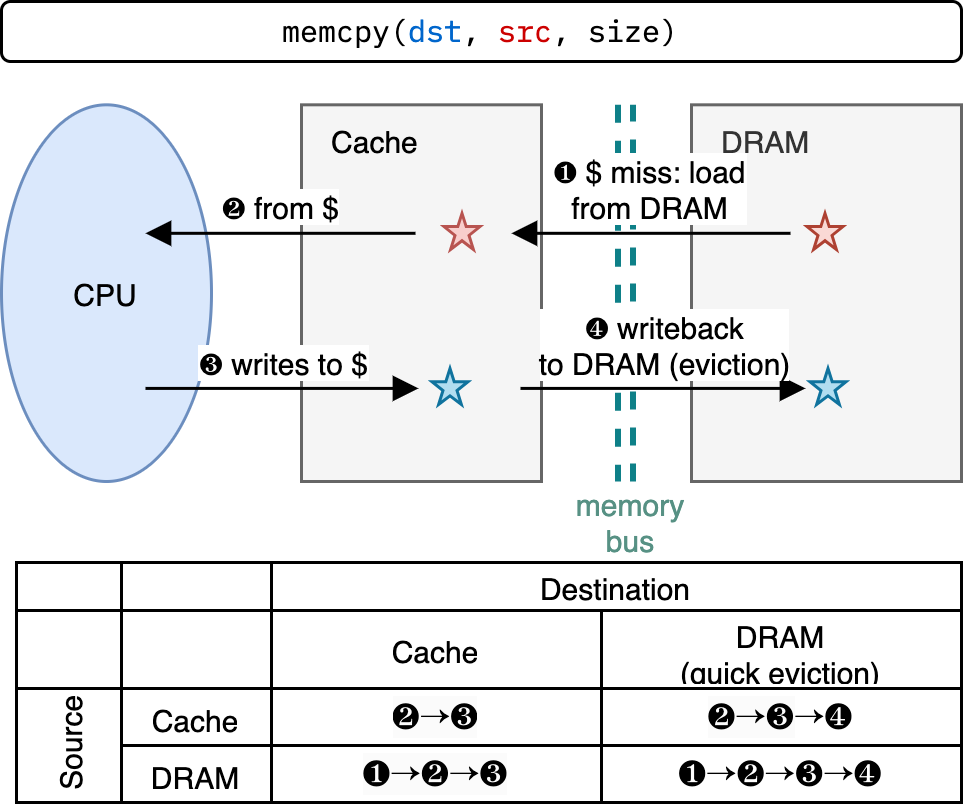}
      \caption{CPU-based memcpy execution path}
      \label{fig:cpu-memcpy}
  \end{subfigure}
  % \hfill
  \hspace{0.1\linewidth}
  \begin{subfigure}[b]{0.4\linewidth}
      \centering
      \includegraphics[width=\linewidth]{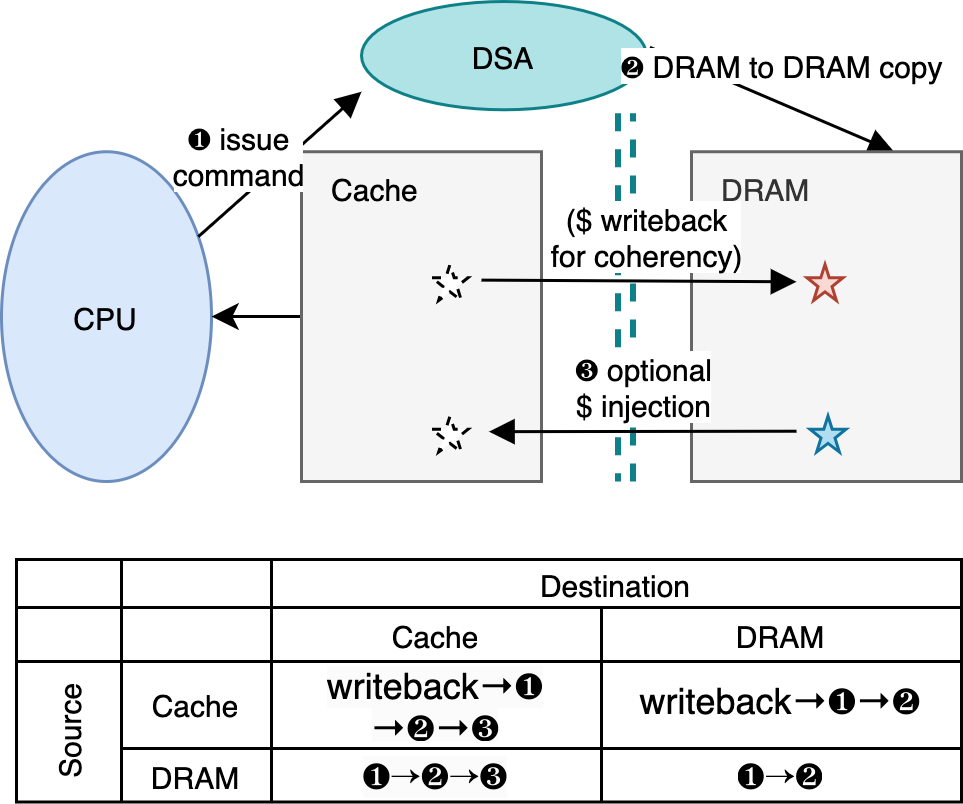}
      \caption{DSA-based memcpy execution path}
      \label{fig:dsa-memcpy}
    \end{subfigure}
  \caption{Comparison of memory copy execution paths between CPU and DSA. CPU-based memcpy naturally integrates with the cache hierarchy, while DSA-based memcpy bypasses the cache, accessing DRAM directly.}
  \vspace{-1.0ex}
  \label{fig:memcpy-path}
\end{figure*}

% \vspace{\baselineskip}
\noindent \textbf{Impact of Page Faults.}\hspace*{0.5em} While DSA supports virtual memory and defers page fault handling to the host, page faults still introduce latency that degrades performance in memory offloading to hardware. As shown in \autoref{fig:pagefaults}, when faults occur, DSA provides no clear advantage over CPU-based \smalltt{memcpy}. In contrast, with pinned memory, DSA significantly outperforms the CPU, even with repeated accesses to the same buffer. These results suggest that frequent page faults can prevent DSA from achieving its theoretical peak performance, emphasizing the importance of pre-mapped or pinned memory regions for effective use.

% \vspace{\baselineskip}
% \noindent
% \textbf{\revblock{Cache Interference Effects.}}\hspace*{0.5em}
% DSA can bypass the CPU cache and perform direct DRAM-to-DRAM copies, which reduces cache pollution. However, if the copied data is accessed immediately by the CPU, bypassing the cache introduces a cold-cache effect and increases memory access latency. 
% To mitigate this, cache injection, a feature that enables data to be routed into the LLC as it is copied, allows selected data to be loaded directly into the last-level cache (LLC) during DSA transfers, reducing latency for immediate post-copy CPU accesses.
% \autoref{fig:cache-injection} shows cache injection improves LLC hit rates in single-threaded execution with prompt reuse and \revblock{without cache line contention}. However, in multi-threaded settings, it increases miss rates due to cache interference. These results suggest that cache injection is only effective under low contention.

% These observations highlight that the benefits of offloading are contingent on workload behavior and synchronization design—factors we explicitly address in our protocol design (\autoref{sec:design}).

\vspace{-1.0ex}
\subsection{Cache Interference and Path Divergence: CPU vs. DSA}

% To illustrate the trade-offs discussed above, we compare CPU-based and DSA-based \smalltt{memcpy} execution paths.
% % In the CPU path (\autoref{fig:cpu-memcpy}), memory copies benefit from cache locality and prefetching, making them suitable for data with immediate reuse. In contrast, the DSA path (\autoref{fig:dsa-memcpy}) avoids cache pollution but may suffer from cold cache effects and coherence overhead if the copied data is reused shortly after.
% CPU-based copies leverage cache locality and prefetching, while DSA bypasses the cache. this trade-off, especially regarding reuse patterns and interference, is discussed in the previous section.
% % memory copy execution paths.
% This contrast highlights the importance of workload-aware offloading strategies. %Depending on the expected reuse pattern and access timing, \sys decides whether to offload memory operations and whether to enable cache injection, as discussed next.
% %
% To illustrate how the trade-offs discussed above manifest in practice, we compare CPU- and DSA-based memory copy execution paths, focusing on their interaction with the memory hierarchy.

Memory offloading alters assumptions about temporal locality in memory hierarchy. Unlike CPU memory operations that benefit from automatic cache retention, memory offloading bypasses caches by default, potentially causing cold-cache effects when data is reused soon after. While this reduces cache pollution, it increases latency for near-term accesses.
To mitigate this, DSA supports explicit cache injection, which routes selected data into the LLC during transfer. As shown in \autoref{fig:cache-injection}, cache injection improves LLC hit rates in single-threaded workloads but can degrade performance under multi-threaded contention.

To illustrate, \autoref{fig:memcpy-path} compares execution paths: CPU copies benefit from cache locality and prefetching, while DSA avoids the cache, risking cold-cache latency. These trade-offs highlight the need for application-level control. Instead of a static, one-size-fits-all configuration, DSA software support should offer a tunable interface, enabling developers to tailor offloading based on reuse patterns, access locality, and IPC integration.

\noindent
\textbf{Execution Implications by Access Direction.}\hspace*{0.5em}We summarize how memory access direction affects DSA offloading decisions.
\begin{tightitemize}
  \item \textbf{Read-In (DRAM → Cache):} CPU loads reused data into cache, leveraging locality. DSA bypasses cache and may cause cold-start penalties (\autoref{fig:cpu-memcpy}). 
  \item \textbf{Write-Out (Cache → DRAM):} DSA avoids polluting cache with write-out data, beneficial for ephemeral data. But reuse-sensitive data may suffer from bypass-induced misses (\autoref{fig:dsa-memcpy}).
\end{tightitemize}

% % \vspace{\baselineskip}
% \noindent
% \textbf{DRAM to Cache (Read-In).}\hspace*{0.5em}
% In CPU-based IPC operations, when data is read into the CPU, it is typically loaded into the cache (\autoref{fig:cpu-memcpy}). This cache residency benefits workloads with high temporal locality, reducing latency for repeated accesses. For such cases, CPU-based memory copying can outperform DSA due to its alignment with existing cache hierarchies.

\begin{table*}[t]
  %\vspace{-2.0ex}
  \centering
  \small
  % \footnotesize
  \caption{Trade-offs of DSA offloading and implications for system design. Each factor highlights a key limitation and how \sys addresses it through configurable or default design decisions.}
  \label{tab:tradeoffs}
  \begin{tabular}{p{2.2cm} p{4.0cm} p{5.3cm} p{4.6cm}}
    \toprule
    \textbf{Factor} & \textbf{Observed Trade-off} & \textbf{Microbenchmark Insight} & \textbf{Design Implications} \\
    \midrule
    \textbf{Data Size} & Offloading not always beneficial for small transfers & Offloading 1MB saves ~33µs; breakeven 4KB raw, higher by setup~\cite{asplos-kuper} & Use CPU-based memcpy for small transfers; apply size-based threshold \\
    \midrule
    \textbf{Page Faults} & DSA supports virtual memory but performance drops with page faults & page faults eliminate DSA speedup; pinned memory yields best performance (\autoref{fig:pagefaults}) & Reuse shared memory to avoid PFs; enforce pre-mapping or pinning \\
    \midrule
    \textbf{Cache Injection} & May improve or harm performance depending on reuse timing & Boosts hit rate in single-threaded case; degrades multi-threaded due to pollution (\autoref{fig:cache-injection}) & Enable cache injection only under low contention. (e.g. single-threaded sync/async modes) \\
    \midrule
    \textbf{Synchronization} & Frequent polling causes bus contention and stalls & \smalltt{UMWAIT} reduces active polling cost but limited to ~25µs (\autoref{fig:polling-strategy}, \cite{intel-dpdk}) & Use hybrid polling (\smalltt{UMWAIT} + timeout); defer checks in pipelined mode \\
    \midrule
    \textbf{Parallelism} & Untapped unless explicitly orchestrated in software & No parallel execution in idxd; DTO blocks until completion & Enforce structured async/pipelined execution to leverage concurrency \\
    \midrule
    \textbf{Software Support} & Existing tools lack intelligent decision logic & idxd offers low-level control; DTO disables parallelism & Provide high-level API with tunable execution modes and cache options \\
    \bottomrule
  \end{tabular}
  \vspace{-2.0ex}
\end{table*}

While DSA offers the potential to accelerate memory operations, its effectiveness depends heavily on workload characteristics and system-level interactions. \autoref{tab:tradeoffs} summarizes key trade-offs identified through microbenchmarks and outlines their design implications for \sys. These insights inform our execution model, which selectively applies memory offloading based on transfer size, thread count, and reuse patterns. \autoref{sec:design} details how these trade-offs affect IPC performance by comparing data movement in CPU- and accelerator-based pipelines.

\section{Design}\label{sec:design}
% Building upon the trade-offs and design considerations identified in the previous sections, we propose \sys, a shared memory-based IPC model that maximizes the benefits of DSA while minimizing its drawbacks.
% \ada{describe what \sys is and connect that to observations from previous section. \sys exposes a suite of IPC implementations that enable the DSA to be leveraged in IPC in a use-case specific manner.... }
% Shared memory-based IPC inherently reuses memory segments, reducing the likelihood of page faults and thereby avoiding the associated overhead. This characteristic makes it particularly well-suited for leveraging DSA efficiently.

To address the challenges in \autoref{sec:tradeoffs}, we present \sys, a runtime that integrates hardware-assited memory offload in IPC suite for intra-node communication over shared memory. \sys supports user-directed offloading, avoiding unnecessary overhead and leveraging DSA when beneficial. Instead of a fixed policy, it provides configurable modes for execution, synchronization, and cache behavior, allowing IPC to align with workload and hardware characteristics. This flexibility enables adaptation to dynamic environments.
% fine-grained adaptation to workload and hardware. % behavior.

\begin{comment}
By reusing shared memory segments, \sys reduces the pagefault overhead. It exposes an API that provides explicit control when offloading is applied, whether cache injection is enabled, and how completions are tracked, %--either synchronously or asynchronously--
depending on application needs.
The following subsections describe the architectural components, execution modes, and supported APIs of \sys.
\end{comment}

\subsection{Overview of the Proposed Design}

\begin{figure*}[t]
    \centering
    \includegraphics[width=0.95\textwidth]{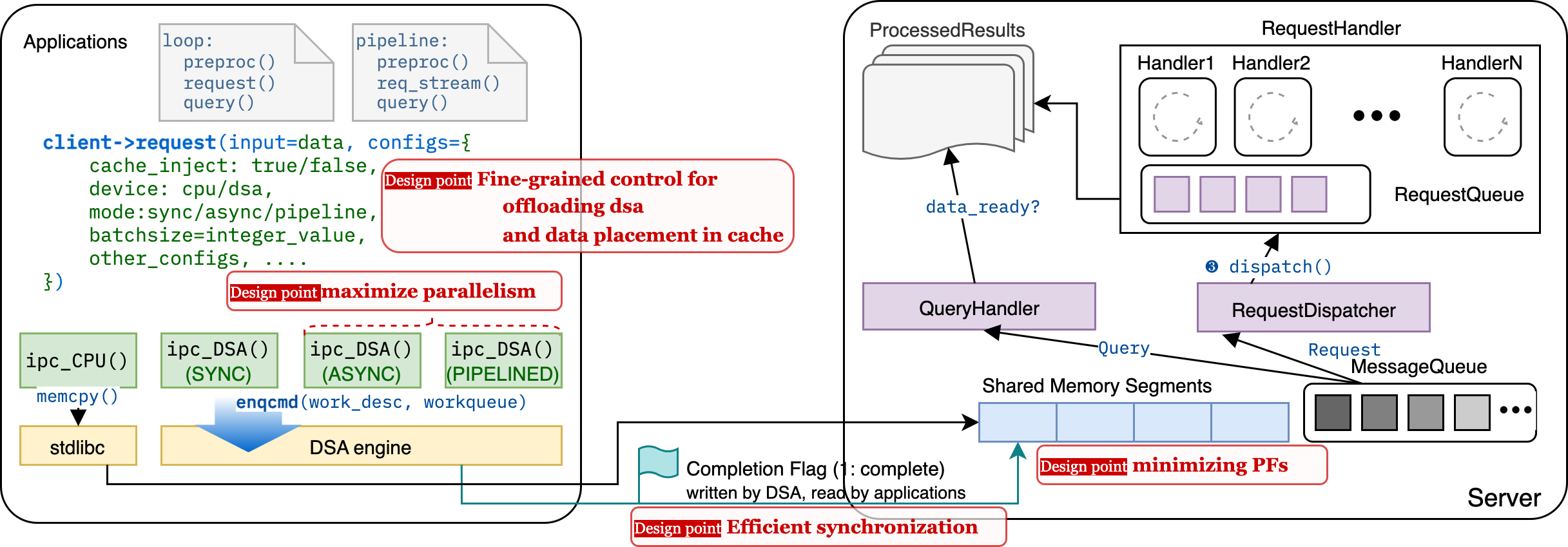}
    \caption{Overview of \sys architecture. Its components reflect key design principles such as page fault avoidance, maximizing parallelism, configurable selective offloading with or without cache injection enabled, and efficient synchronization.}
    \vspace{-2.5ex}
    \label{fig:design}
\end{figure*}

The overall execution flow and system architecture are shown in \autoref{fig:design}. \sys is a shared memory-based IPC suite that supports multi-client connections and selective offload of data movement to the DSA. The server comprises a message queue, DSA engine, request dispatcher, request handler, and query handler, enabling asynchronous batching and efficient CPU-DSA overlap.

% \begin{figure}[h]
%     \centering
%     \includegraphics[width=0.9\columnwidth]{figures/design.png}
%     \caption{System architecture of \sys. Components reflect key design principles such as page fault avoidance, selective offloading, and parallel execution.    \ada{this figure and associated description is too low level, let's discuss how to abstract some of the more implementation detail from higher level design components}}

%     \label{fig:design}
% \end{figure}
% \begin{figure}[h]
%     \centering
%     \includegraphics[width=1.0\columnwidth]{figures/design2.pdf}
%     \caption{System architecture of \sys. Components reflect key design principles such as page fault avoidance, selective offloading, and parallel execution.    \misun{is this figure better? I know it isn't the pretties, but if having those design principles might help i can prettify this. My idea is having figure star and make the flow horizontal}}

%     \label{fig:design}
% \end{figure}

The system design is guided by a set of hardware-aware principles derived from the trade-offs discussed in \autoref{sec:tradeoffs}:

% \begin{tightitemize}
%   \item \textbf{Avoiding Page Fault Overhead.}  
%   To minimize the impact of virtual memory translation and remapping, \sys reuses shared memory segments across transfers. This is realized through the persistent \textit{message queue} and \textit{shared memory allocation strategy}, which maintain buffer continuity.

\noindent
\textbf{(1) Shared Memory Reuse.}\hspace*{0.5em}
To reduce remapping overhead and avoid page faults, \sys reuses shared memory segments across transfers. This is enabled by a persistent \textit{message queue} and pre-allocated \textit{shared memory} to maintain buffer continuity.

\noindent
\textbf{(2) Flexible Transfer Offloading.}\hspace*{0.5em}
\sys exposes an interface for tuning offload decisions based on data size and workload characteristics. Unlike static offload frameworks (e.g. Intel DTO), which statically offloads all transfers, \sys supports adaptive strategies, allowing users to balance overhead and latency.

\noindent
\textbf{(3) Built-in Support for CPU-DSA Parallelism.}\hspace*{0.5em}
\sys supports asynchronous transfers by integrating synchronization and coordination mechanisms typically left to the user in low-level APIs. This enables concurrent CPU-side execution while offloading data movement to DSA, a capability not supported by synchronous-only baselines like DTO.

\noindent
\textbf{(4) Cache-Aware Data Placement.}\hspace*{0.5em}
\sys exposes cache injection as an API knob. In multi-client settings, the server shares execution context so clients can enable injection selectively, based on reuse likelihood and system load.

\noindent
\textbf{(5) Reducing Synchronization Overhead.}\hspace*{0.5em}
\sys avoids frequent polling in async modes by deferring completion checks. The \textit{query handler} uses deterministic time prediction and \smalltt{UMWAIT}-based waiting for low-latency coordination.

% \end{tightitemize}

% On the client side, a dedicated DSA engine handles data transmission, allowing fine-grained control over the type of device used for data movement, its operation mode (synchronous, asynchronous, or pipelined), and whether cache injection is enabled (as depicted in the table within the figure). Furthermore, beyond simple loop-based execution, the programming model explicitly enforces pipelining, ensuring that workloads with high throughput demands can fully exploit parallel execution to maximize performance.

% To achieve optimal performance, our design follows several key principles. First, parallelism must be maximized through aggressive offloading, ensuring that DSA operations and CPU computations overlap as much as possible. Second, our programming model naturally accounts for data usage prediction, allowing the system to determine whether data should be cached or bypassed based on its expected access pattern. Lastly, to ensure flexibility and adaptability, we provide a configurable API that allows seamless switching between CPU and DSA execution, cache injection on or off, and synchronous or asynchronous modes. The following sections elaborate on how these design principles translate into specific design decisions in our proposed model.

Together, these principles are embodied in a modular IPC stack that can be adapted to varying workloads and system-level behaviors. %The following sections describe each optimization strategy in more detail and explain how they are exposed through a unified API.

\subsection{IPC API Specification}\label{subsec:api}
% To abstract the complexities of memory offloading and cache behavior,
\sys exposes a set of configurable IPC APIs %. These APIs are
designed to offer flexibility across diverse workloads while maintaining performance portability and ease of use. The interface supports multiple execution modes and allows explicit user control over data movement, synchronization, and cache injection policies.

\begin{figure}
    \flushleft
    \begin{subfigure}[b]{0.90\linewidth}
        \includegraphics[height=1.5cm]{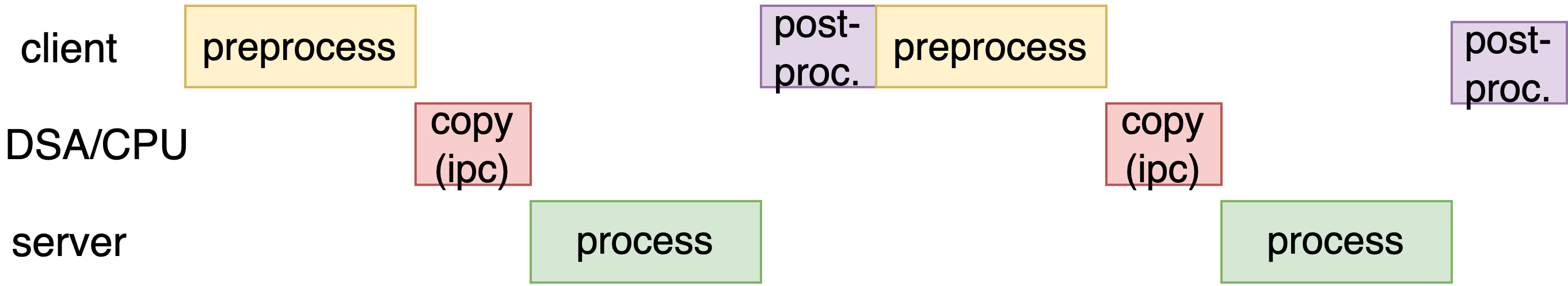}
        \caption{2 passes of synchronous execution}
        \label{fig:pl-sync}
    \end{subfigure}
    \begin{subfigure}[b]{0.90\linewidth}
        \includegraphics[height=1.5cm]{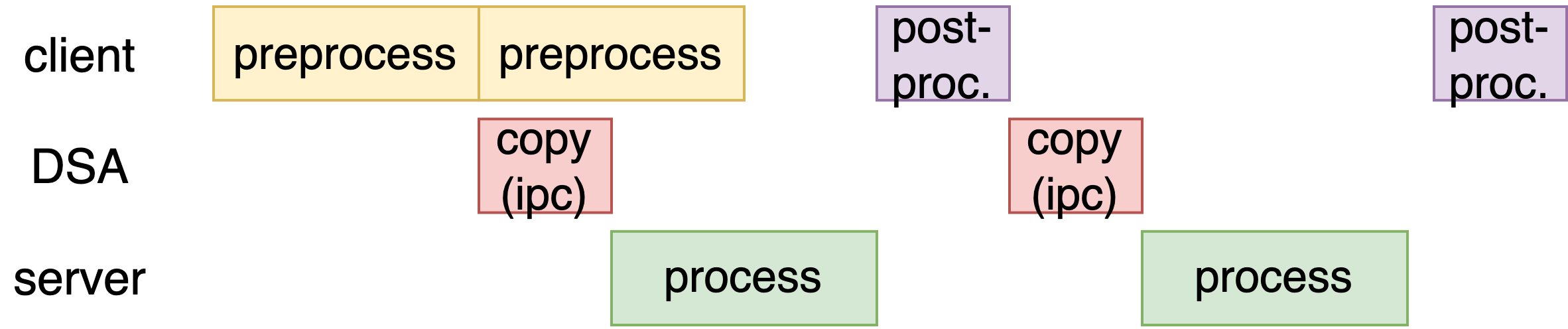}
        \caption{2 passes of asynchronous execution}
        \label{fig:pl-async}
    \end{subfigure}
    \begin{subfigure}[b]{0.90\linewidth}
        \includegraphics[height=1.5cm]{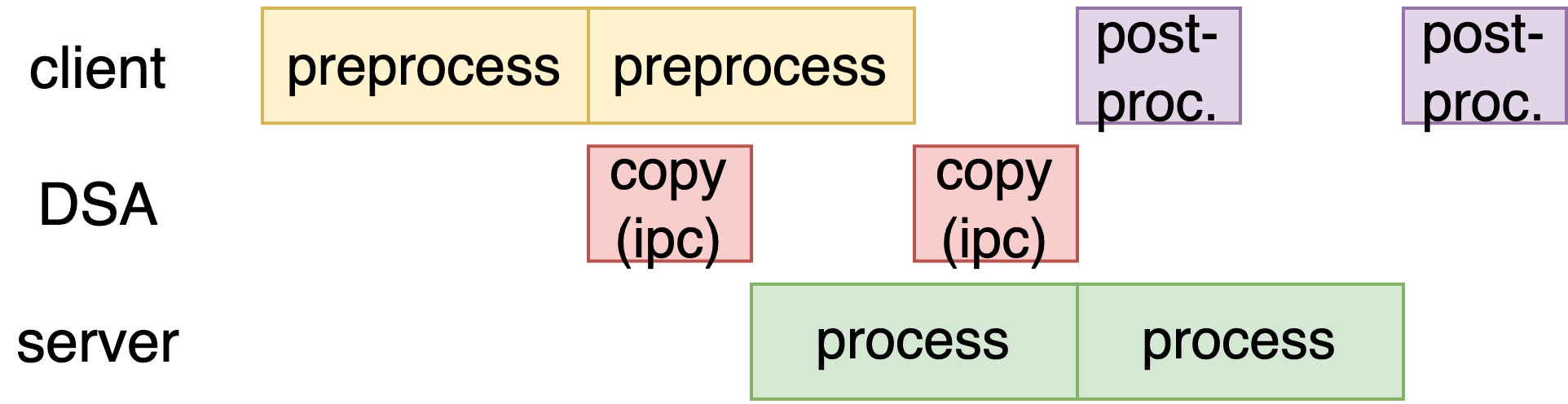}
        \caption{2 passes of pipelined execution}
        \label{fig:pl-pl}
    \end{subfigure}
    \caption{Execution mode structure in \sys. Each mode differs in synchronization and overlap strategy.}
    \vspace{-5.0ex}
    \label{fig:pipeline}
\end{figure}

\noindent
\textbf{Execution Modes.}\hspace*{0.5em}
\sys supports three execution modes -- \smalltt{sync}, \smalltt{async}, and \smalltt{pipelined} (\autoref{fig:pipeline}). Each mode offers different trade-offs between memory offloading, synchronization, and CPU-DSA overlap. Users can choose the appropriate mode based on workload characteristics and performance goals. %Their structure is illustrated in %\autoref{fig:pipeline}.

\begin{tightitemize}
    \item \textbf{Synchronous Mode:} \smalltt{sync} mode (\autoref{fig:pl-sync}) executes memory operations in a blocking manner. The issuing thread stalls until the operation completes, making it suitable for latency-sensitive, sequential tasks. Cache injection may be enabled to reduce cold-cache penalties in settings with low cache contention.

    \item \textbf{Asynchronous Mode:} \smalltt{async} mode (\autoref{fig:pl-async}) decouples submission from completion. Requests return immediately, and completion must be checked explicitly. This allows the CPU to perform useful work while the operation is in-flight, making it appropriate for moderately parallel workloads or where data transfer latency can be hidden.

    \item \textbf{Pipelined Mode:} \smalltt{pipelined} mode (\autoref{fig:pl-pl}) issues memory operations in batches to improve throughput and reduce overhead. It defers individual completion checks and reuses buffers across stages. This mode is effective when the workload involves large memory traffic and when preprocessing can be overlapped. %However, it
      It may underperform if preprocessing dominates execution.
\end{tightitemize}

% Detailed interactions of these modes with synchronization and completion tracking are further described in \autoref{subsec:design-internals}.

\noindent
\textbf{Configurable Parameters.}\hspace*{0.5em}
%The core APIs are centered around a small number of configurable entry points.
All modes support the same functional interface, with additional arguments that configure execution behavior. To support workload-specific tuning, the API exposes several parameters:

\begin{tightitemize}
\item \textbf{Offload control:} Applications can choose whether to use DSA or CPU for a particular memory operation. This allows bypassing DSA for small or non-beneficial transfers.
\item \textbf{Cache injection policy:} The user may explicitly enable or disable cache injection. By default, \sys applies mode-specific policies based on empirical performance trends. %, e.g., enabling injection in synchronous mode and disabling it in pipelined mode. %These defaults are detailed in \autoref{subsec:design-internals}.
\item \textbf{Execution mode:} %The user can select between synchronous, asynchronous, or pipelined execution. This allows
  Applications can choose the most suitable execution model based on their performance requirements.
\end{tightitemize}

These parameters offer fine-grained control while maintaining sensible defaults, enabling users to select the most appropriate memory offloading strategy as needed. 

\subsection{Design Internals}\label{subsec:design-internals}
% To realize the flexibility and performance guarantees of the API described in \autoref{subsec:api}, \sys employs a modular and configurable backend architecture. This section details the internal organization of the system, focusing on shared memory management, execution engines, and request handling strategies across different modes.
% Next, we provide additional detail on the internals of \sys. 

\noindent
\textbf{Shared memory region reuse.}\hspace*{0.5em}
\sys uses persistent shared memory regions to minimize page fault overhead. At connection setup, the server allocates a fixed-size memory pool and assigns each client a dedicated queue pair, transmit (client-to-server) and receive (server-to-client) buffers, mapped once and reused throughout the session. This eliminates remapping costs and ensures stable, low-latency memory access, enabling efficient DSA transfers. The design is inspired by RDMA queue pairs but tailored to DSA-based copy semantics.

\noindent
\textbf{Asynchronous DSA Engine.}\hspace*{0.5em}
To enable parallelism and hide memory latency, \sys includes a lightweight asynchronous engine for managing DSA command dispatch and completion. It abstracts low-level DSA primitives and provides a clean interface to the IPC driver.
Requests are routed to mode-specific paths, where the engine handles command issuance, completion tracking, and batching in \smalltt{pipelined} mode. A hybrid polling strategy balances low CPU overhead with responsive completion checks.

\noindent
\textbf{Completion Tracking and Hybrid Polling Strategy.}\hspace*{0.5em}
% Polling introduces both explicit and implicit performance costs. Explicitly, busy polling consumes CPU cycles by keeping the core occupied while waiting for operation completion. More critically, it also introduces a range of implicit system-level penalties~\cite{polling-mem-bandwidth, pipeline-stalls}. Polling on uncacheable device memory can monopolize the memory bus, causing contention even for single-byte reads. It may also trigger strict memory ordering constraints that block out-of-order execution and stall pipelines. Furthermore, implicit memory fences and cache invalidation events can degrade cache efficiency and introduce hidden performance variability across microarchitectures.
% 
% Polling frequency introduces another trade-off between latency and CPU efficiency.
Frequent polling enables prompt completion detection but increases overhead. In contrast, infrequent polling lowers overhead at the risk of delayed response.
%The trade-offs between polling strategies—specifically in terms of latency and CPU overhead—are already illustrated in \autoref{fig:polling-strategy}, which quantifies their behavior for a 1MB transfer.
% Pros and cons of user-mode synchronization mechanisms are summarized in \autoref{tab:polling-cost}. \ada{reference figure in earlier section, removing table since redundant info  We therefore omit a separate summary table.}
\begin{comment}
  
\begin{table}[h]
    \caption{User-mode polling mechanisms and their limitations.}
    \label{tab:polling-cost}
    \begin{tabular}{c c}
        \toprule
        Methods & Downsides \\
        \midrule
        \begin{tabular}[l]{@{}l@{}}Read and yield \\ (\smalltt{sched\_yield()})
        \end{tabular} &  
        \begin{tabular}[l]{@{}l@{}} 
        Too frequently reading 
        \end{tabular} \\
        \midrule
        Read and sleep &  
        \begin{tabular}[l]{@{}l@{}} 
        Too infrequently reading
        \end{tabular} \\
        \midrule
        \smalltt{UMONITOR} and \smalltt{UMWAIT} &  
        \begin{tabular}[l]{@{}l@{}} 
        Unable to wait longer \\ than 100k cycles ($\sim 25\mu s$)~\cite{intel-dpdk}
        \end{tabular} \\
        \bottomrule
    \end{tabular}
    \end{table}

  \end{comment}  
  To reconcile these trade-offs, \sys implements a hybrid polling strategy that combines time-based deferral with passive wait instructions.
 % Upon initiating a DSA operation, the system estimates the expected completion time based on the data size. Polling is deferred until this estimated window, at which point \smalltt{UMWAIT} is used to passively monitor the completion flag without consuming active CPU cycles. 
% Several polling techniques are available in user space, each presenting distinct trade-offs in terms of latency, CPU overhead, and responsiveness. \autoref{tab:polling-cost} summarizes the characteristics and limitations of these approaches.
%
  While \smalltt{UMWAIT} minimizes CPU involvement, it is limited to a 100K cycle ($\approx$25$\mu$s) wait period~\cite{intel-dpdk}.
  %\ada{may remove fig for space} As shown in \autoref{fig:latency-per-size}, DSA copy latency grows linearly with transfer size (~33.4$\mu$s/MB), making simple \smalltt{UMWAIT}-only approaches insufficient for large offloads. %To address this, \sys computes a deferral interval based on size, sleeping initially and invoking passive polling only near the expected completion. 
%

\begin{figure}[t]
   \centering
   % \hspace{-2.5em}
   \includegraphics[width=0.95\columnwidth]{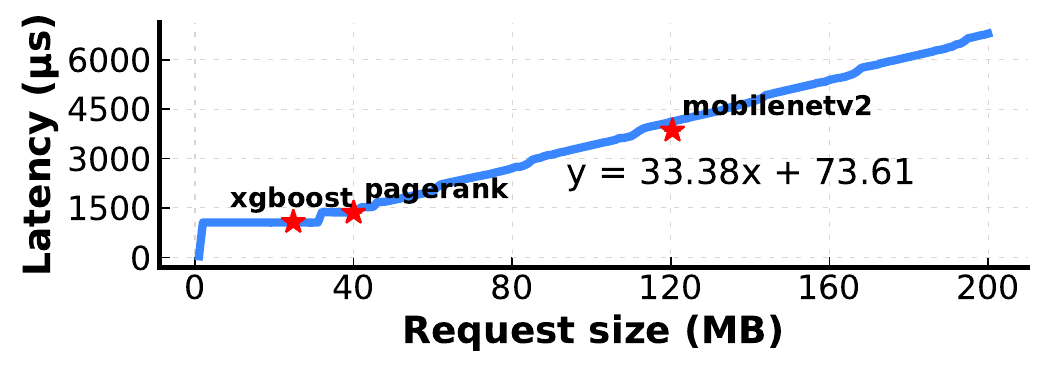}
   \vspace{-1.0ex}
   \caption{The latency of \smalltt{memcpy} increases linearly with the target buffer size, at approximately 33.4~$\mu$s per 1MB on our hardware.} 
   \vspace{-3.5ex}
   % (Experimental setup: within CPU socket, 1 DSA engine, 1 work queue, 1 batch size.}
   \label{fig:comp-track}
\end{figure}

  %To address this,
  Instead, \sys implements a size-aware deferral mechanism that estimates the expected completion time based on the request data size and delays polling accordingly. 
  % \ada{needs some additional clarification. Are the two parameters hardware/platform-specific or workload specific too? Can you comment how diff is it to determine them/can this be part of \sys as some script? even if needs to be done on per workload basis. } 
  Let $L$ be the estimated latency in microseconds: $L = L_{\text{fixed}} + \alpha \cdot \smalltt{size\_in\_MB}.$ Both $L_{\text{fixed}}$ and $\alpha$ are machine-dependent but remain consistent across workloads for a given system (\autoref{fig:comp-track}). \sys includes a profiling script that automatically derives these parameters during the initial deployment. 
  % \misun{How about this?}\ada{this is ok. may need to remove figure due to space}
  % ; this one-time procedure enables reuse of the calibrated values for all subsequent workloads on the same machine.

%
%\[
%L = L_{\text{fixed}} + \alpha \cdot \smalltt{size\_in\_MB}
%\]
%

In our implementation, we use $L_{\text{fixed}} = 73.6\mu s$ and $\alpha = 33.4~\mu s/\text{MB}$, based on empirical measurements. We repeated 100 latency measurements with varied $L_{fixed}$ and $\alpha$ (std. dev. $<$ 2\%). The copy latency was highly consistent since both the instruction path and the memcpy engine’s transfer function are hardware-defined. These parameters may vary across machines but remain consistent within the same system. A helper script in the source automatically recalibrates them per node for reproducibility.
% (\autoref{fig:latency-per-size}). %The polling flow is as follows:
% These values are derived from 100 repeated measurements, with the maximum deviation from the mean observed to be less than 0.01~$\mu$s.
\sys computes $L$ based on transfer size, and sleeps for $L \cdot 0.95$ to yield the CPU. It then begins polling using \smalltt{UMWAIT}.
\begin{comment}
\begin{tightenumerate}
  \item Compute $L$ based on transfer size.
  \item Sleep for $L \cdot 0.95$ to yield the CPU.
  \item Begin polling using \smalltt{UMWAIT}.
  \item If the flag is not yet complete, repeat or fall back to yield-based polling.
  \end{tightenumerate}
  \end{comment}
This hybrid strategy reduces unnecessary polling by deferring checks until completion is likely, based on size-aware latency prediction. Once polling begins, \smalltt{UMWAIT} enables passive waiting without CPU burn. This approach balances latency and efficiency, avoiding the cost of busy-loop polling while preserving timely completion detection.
It also enables adaptive runtimes to adjust polling behavior dynamically based on request size and system load, navigating the trade-off between responsiveness and efficiency.

\noindent
\textbf{Request Batching Support.}\hspace*{0.5em}
To enable high-throughput, especially in \smalltt{pipeline} mode, \sys supports application-level request batching. Incoming messages are classified as either requests or result queries. Requests are dispatched to workload-specific handlers (e.g., \smalltt{MobileNetV2}, graph processing), which are registered via a unified interface. Handlers execute asynchronously and write results to shared memory.
By decoupling submission from completion, \sys allows deferred result collection and batch processing. This improves buffer reuse and reduces synchronization overhead, enabling multiple outstanding requests to be processed collectively with minimal coordination.

\noindent
\textbf{Execution Stack Internals.}\hspace*{0.5em}
As shown in \autoref{fig:design}, \sys separates request submission, execution, and completion tracking across distinct components.
The \textit{RequestDispatcher} receives incoming requests from a message queue and routes them to \textit{RequestHandlers}, which are instantiated per thread context. Handlers issue DSA commands using mode-specific logic (e.g., \smalltt{sync}, \smalltt{async}, \smalltt{pipelined}). In \smalltt{pipelined} mode, requests are batched to maximize throughput and amortize overhead.
The \textit{QueryHandler} tracks completion by polling shared memory result flags. It can be invoked explicitly in \smalltt{pipelined} mode. By decoupling completion tracking from request handling, \sys enables configurable synchronization and supports hybrid polling via user-mode interrupt (\smalltt{UMWAIT}).
Together, these components allow \sys to adapt execution to workload requirements while isolating low-level control from application-facing APIs.

\section{Implementation}\label{sec:impl}
% What programming languages is it built with?
% How many lines of code, and where can it be found?

% 5642 + 6158 lines
\sys is implemented in C++17 and consists of approximately 12,000 lines of code.
% Its modular design separates core components from auxiliary utilities to improve maintainability.
The full source code, along with build instructions, is publicly available on GitHub to support reproducibility and community contributions.
The system is built on the Intel IDXD driver, using its latest version to interface with the DSA hardware. It provides a high-level API over IDXD to simplify DSA usage while preserving performance. The \smalltt{accel-config} library is used to configure DSA devices.
The implementation leverages Intel-specific instructions such as \smalltt{UMONITOR}, \smalltt{UMWAIT}, and \smalltt{ENQCMD}, along with intrinsics for efficient execution. Standard functionality is implemented using C++17, System V IPC, and POSIX libraries. ONNX~\cite{onnx}, XGBoost~\cite{xgboost}, BoostGL~\cite{boostgl}, MilvusDB~\cite{milvusdb}, and OpenCV are used in benchmark workloads.

\begin{comment}
This section outlines the key components and interactions in the implementation.

\subsection{\sys APIs in Action}
\sys is designed to be user-friendly, providing a high-level API that abstracts the complexities of DSA integration. The API is structured to facilitate seamless interaction with the underlying hardware while allowing users to customize execution modes, cache management, and synchronization strategies.
% This design ensures that developers can easily leverage DSA's capabilities without delving into low-level implementation details.

\noindent \textbf{Parallel Execution: Enforcing Overlap.}\hspace*{0.5em}
Each design principle in \smalltt{\sys} is directly reflected in its API, ensuring that parallel execution, cache management, and polling efficiency are seamlessly integrated into the system.
The API explicitly differentiates execution modes, ensuring that DSA operations and CPU execution naturally overlap without requiring manual tuning. \autoref{lst:api-example} demonstrates how the API achieves this:
\end{comment}

% \vspace{-2.0ex}
\begin{tiny}
% \begin{lstlisting}[language=C++, label={lst:api-example}, caption={\sys API for parallel execution modes}]
\begin{lstlisting}[language=C++, basicstyle=\footnotesize\ttfamily, label={lst:api-example}, caption={\sys API for parallel execution modes}]
  // Default: Synchronous execution (blocking)
  client->request(mode="sync", op="mobilenetv2", data);
  // Asynchronous: Non-blocking, requires explicit completion check
  future = client->request(mode="async", op="mobilenetv2", data);
  future.get();  // Waits for completion

  // Pipeline: Batches multiple requests, polling handled at batch level
  // In the real implementation, it is encapsulated in a dedicated function
  for (int i = 0; i < batch_size; i++) {
    job_ids[i] = client->request(mode="pipeline", op="mobilenetv2", data[i]);
  }
  // doing other things\dots
  for (int i = 0; i < batch_size; i++) {
    // Wait for all jobs to complete
    results[i] = client->query(job_ids[i]);
  }
\end{lstlisting}
\end{tiny}

\noindent
\textbf{\sys API.}\hspace*{0.5em} \autoref{lst:api-example} illustrates the \sys API across three execution modes.
Each request specifies an operation (e.g., \smalltt{mobilenetv2}) via \smalltt{op} and provides input through \smalltt{data}.
\smalltt{sync} mode blocks until completion, resembling \smalltt{memcpy}.
\smalltt{async} mode returns a future for non-blocking execution, with explicit synchronization via \smalltt{get()}.
\smalltt{pipeline} mode queues requests internally and batches them for processing, returning job IDs for later result queries. This defers polling and reduces overhead.
The server includes concurrency metadata to help clients adapt cache injection. By default, injection is enabled in sync mode, conditionally enabled in async (if single-threaded), and disabled in pipeline mode due to decoupled execution.

\noindent
\textbf{Configuration Effort. }
\sys exposes three key parameters: \smalltt{device} (\smalltt{cpu}, \smalltt{dsa}), \smalltt{cache\_injection} (\smalltt{on}, \smalltt{off}), and execution \smalltt{mode} (\smalltt{sync}, \smalltt{async}, \smalltt{pipeline}). Among these, \smalltt{mode} is compile-time fixed, as it dictates control flow. For endpoints with downstream processing, \smalltt{async} or \smalltt{pipeline} generally outperform \smalltt{sync}, with the former favoring latency and the latter throughput.
\smalltt{device} selection depends on transfer size. Real applications tend to favor \smalltt{dsa} at larger sizes than microbenchmarks suggest~\cite{asplos-kuper}, due to preprocessing, cache traffic, and CPU-DSA bus contention.
\smalltt{cache\_injection} is effective when data is immediately reused, but should be avoided under heavy cache contention, where it can cause pollution and invalidation. \sys uses a heuristic default -- enabled for \smalltt{sync}/\smalltt{async}, disabled for \smalltt{pipeline} -- but allows user overrides.

\section{Evaluation}\label{sec:eval}

%\autoref{sec:tradeoffs} identified the fundamental limitations of naive DSA integration through targeted microbenchmarks. These system-level analyses guided the design of \sys in \autoref{sec:design}, which introduces mechanisms such as mode-aware offloading, cache injection control, and hybrid polling.

%In this section, we evaluate whether these design choices result in measurable end-to-end performance benefits across representative workloads. Rather than presenting additional synthetic benchmarks, we assess \sys's effectiveness in realistic inference and analytics pipelines where memory movement is a key bottleneck.

% The demonstrate experimentally that \sys provides opportunities for end-to-end throughput and latency improvements using two application benchmarks. %, with more flexible programming models than existing DSA-based IPC solutions.
In this section, we experimentally demonstrate that \sys improves end-to-end throughput and latency across representative application benchmarks
% , while enabling more flexible programming models than existing DSA-based IPC solutions.
% \ada{is there data that justifies the flexibility it provides for different APIs?} 
%
% In addition, using low-level data, we
and demonstrate that the design choices in \sys are effective in addressing the data-movement related hardware inefficiencies 
%\ada{still needs a pass to explain where precisely the results show that. if not, remove those results.}
%\ada{any additional data to say the implementation is good/lightweight... ?}

\subsection{Experimental Methodology}
\noindent{\bf Testbed.} The experimental platform is summarized in \autoref{tab:experiment-setup}. We use a system equipped with an Intel Sapphire Rapids processor , an NVIDIA A100 GPU and a single DSA device. 

\begin{table}[h!]
  \centering
  \footnotesize
  \caption{Experimental Setup}
  \label{tab:experiment-setup}
  \vspace{-1.0ex}
  \begin{tabular}{
    p{1.3cm}  % Component
    p{5.2cm} % Specification
  }
    \toprule
    \textbf{Component} & \textbf{Specification} \\
    \midrule
    CPU & Intel(R) Xeon(R) Gold 6438Y+ 4.0GHz \\
    & (Sapphire Rapids), 32 cores \\
    \midrule
    GPU & NVIDIA A100 (PCIe, 40GB HBM2 memory, 6912 CUDA cores, 312 TFLOPS FP16)\\
    \midrule
    Cache & 60MiB LLC \\
    \midrule
    DSA & 1 Intel DSA device with 1 workqueue \\
    \midrule
    RAM & 704GB DDR4 4800MT/s \\
    \midrule
    OS & Ubuntu 22.04.5 LTS, Kernel 6.5.0-41 \\
    \midrule
    Compiler & GCC 11.4.0 \\
    \midrule
    Libraries & glibc 2.35, accel-config, numactl, DTO, \\
    & PyTorch 2.1.2, ONNX (1.19.1), OpenCV, \\
    & XGBoost, BoostGL, OpenVINO \\
    \bottomrule
  \end{tabular}
\end{table}

\noindent{\bf Workloads. }  
% \ada{The evaluation uses two application benchmarks, ... sumarized in Tables ... make it clear what constitutes a single request in these cases.}
The evaluation uses five application benchmarks, MobileNetV2, XGBoost, PageRank, MilvusDB and Vision Transformer (ViT) summarized in \autoref{tab:combined-pipeline}.
The selected benchmarks collectively capture complementary dimensions of data movement stress. MobileNetV2 models dense tensor streaming typical of inference; XGBoost highlights fine-grained feature batching; PageRank exposes irregular, reuse-heavy graph traversal; MilvusDB reflects batched query execution in modern data stores; and ViT represents a modern deep learning GPU-intensive workload. Together, they cover dense, sparse, and batched access patterns --- the major sources of IPC cost in data-intensive systems, providing a balanced coverage.

% \smalltt{MobileNetV2} represents a video analytics pipeline, where each client performs basic preprocessing on image frames (e.g., resizing and color normalization) before transmitting them to a service for convolutional neural network (CNN) inference. The resulting labels are then postprocessed and visualized. This workload exhibits a relatively balanced distribution of computation across the preprocessing, inference, and postprocessing stages. Each request in MobileNetV2 corresponds to a batch of 80 preprocessed images (160MB total).
% \smalltt{XGBoost} represents a predictive analytics workload, in which the client extracts features from tabular input data (e.g., medical records) and sends them to a server for inference. Due to the high dimensionality and complexity of the feature transformation pipeline, the preprocessing stage dominates the overall computation. In this scenario, offloading memory transfers via DSA plays a key role in reducing client-side latency. A single request in XGBoost consists of 200{,}000 tabular entries from the breast cancer dataset (~25MB). These benchmarks represent common client-server inference pipelines with differing pipeline structures and memory footprints.

\begin{table}[h!]
  \centering
  \footnotesize
% \begin{subtable}[t]{0.48\textwidth}
  \centering
  \caption{Pipeline stages per benchmark workload.}
  \label{tab:pipeline-detail}
  \begin{tabular}{
    p{1.40cm}  % Benchmark
    p{1.84cm}  % Pre-processing
    p{2.05cm}  % Processing
    p{1.87cm}  % Post-processing
  }
    \toprule
    \textbf{Benchmark} & \textbf{Pre-processing} & \textbf{Processing} & \textbf{Post-processing} \\
    \midrule
    MobileNetV2 &
      Image decoding, resizing, normalization &
      CNN inference using ONNX Runtime &
      Parsing, formatting output \\
    \midrule
    XGBoost &
      % Reading structured data, feature vector construction &
      Building feature vector &
      Inference using pre-trained boosted trees &
      Parsing prediction output \\
    \midrule
      PageRank &
      Building adjacency list &
      % Parsing input graph files, adjacency list construction &
      Iterative PageRank computation &
      % Extracting top-10 vertices and displaying output \\
      Extracting top-10 vertices \\
    \midrule
    MilvusDB w/ image embeddings &
    % Image decoding, resizing, embedding extraction via CNN &
    Image decoding, resizing, embedding extraction &
    Approximate NN, top-3 most similar images &
    % Approximate NN, formatting top-3 most similar images &
    Parsing prediction output \\
    \midrule
      Vision Transformer (ViT) &
      Image decoding, resizing, normalization &
      Deep learning transformer-based image inference &
      Parsing, formatting output
    \\
    \bottomrule
    
  \end{tabular}
  \vspace{-2.0ex}
% \end{subtable}
%   \hfill
% \begin{subtable}[t]{0.48\textwidth}
%   \centering
%   \caption{Benchmark summary and pipeline characteristics.}
%   \label{tab:benchmark-summary}
%   \begin{tabular}{
%     p{1.2cm}  % Benchmark
%     p{2.0cm}  % Pipeline Balance
%     p{1.5cm}  % Size
%     p{2.3cm}  % Note
%   }
%     \toprule
%     \textbf{Benchmark} & \textbf{Pipeline balance} & \textbf{size(req) /size(resp)} & \textbf{Note} \\
%     \midrule
%     {MobileNetV2} &
%       Balanced &
%       160MB / 320KB &
%       Large input, compute-heavy \\
%     \midrule
%     {XGBoost} &
%       Preprocess-heavy &
%       25MB / 800KB &
%       Host-bound bottleneck, limited memory reuse \\
%     \midrule
%       PageRank &
%       Process-heavy &
%       76MB / 320KB &
%       Memory bound, frequent memory reuse \\
%     \midrule
%       MilvusDB &
%       Postprocess-heavy (relatively) &
%       1MB / 345MB &
%       Bulky response \\
%     \bottomrule
%   \end{tabular}
% \end{subtable}
  % \caption{
    % Client-server application pipelines used to evaluate DSA-accelerated memory movement. %These pipelines emulate scenarios such as edge inference, online model serving,
    % , and graph processing,
    %with different characteristics in terms of pipeline stage balance, data volume, and system bottlenecks.
  % }
  \label{tab:combined-pipeline}
\end{table}

\begin{figure*}[h!]
  \centering
  \includegraphics[width=0.8\linewidth]{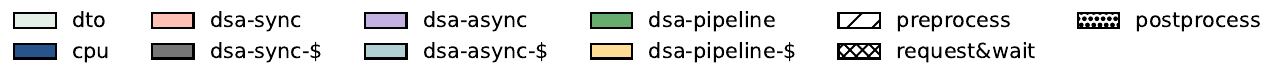}
  \includegraphics[width=\linewidth]{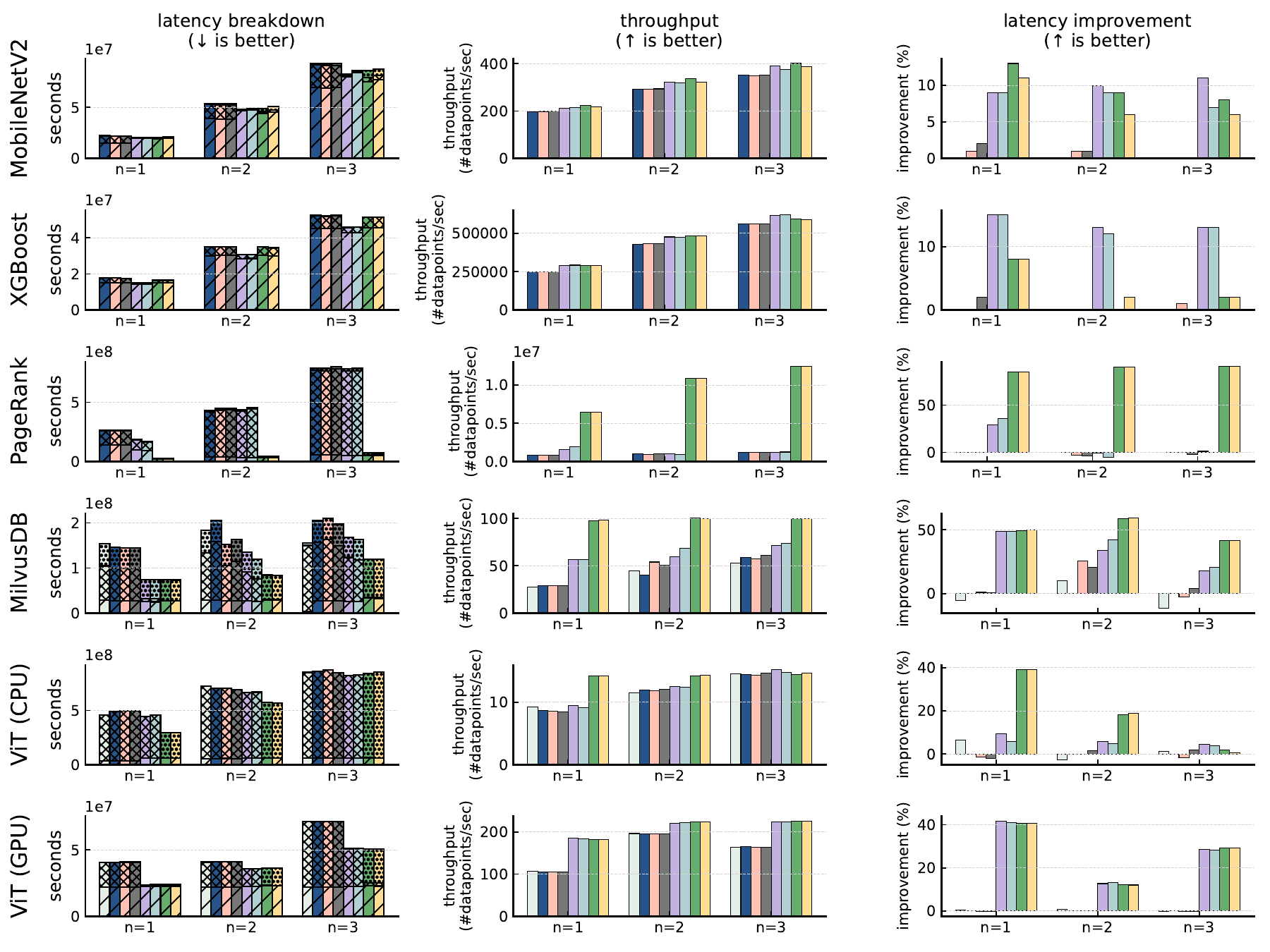}
  % \begin{subfigure}[b]{0.33\linewidth}
  %     \centering
  %     \includegraphics[width=\linewidth]{figures/01_latency-breakdown_n=1.pdf}
  %     \caption{n=1, \textit{undersubscribed}}
  %     \label{fig:main-n1}
  % \end{subfigure}
  % \hfill
  % \begin{subfigure}[b]{0.33\linewidth}
  %     \centering
  %     \includegraphics[width=\linewidth]{figures/01_latency-breakdown_n=2.pdf}
  %     \caption{n=2, \textit{matched}}
  %     \label{fig:main-n2}
  % \end{subfigure}
  % \hfill
  % \begin{subfigure}[b]{0.33\linewidth}
  %     \centering
  %     \includegraphics[width=\linewidth]{figures/01_latency-breakdown_n=3.pdf}
  %     \caption{n=3, \textit{oversubscribed}}
  %     \label{fig:main-n3}
  % \end{subfigure}
  \caption{
    Impact on execution time breakdown (left), throughput in images/sec (middle), and end-to-end latency improvement over CPU-based baseline (right) across for different \sys IPC implementations for \smalltt{MobileNetV2}, \smalltt{XGBoost}, \smalltt{PageRank}, \smalltt{MilvusDB},
    \smalltt{ViT (CPU)},
    \smalltt{ViT (GPU)} under varying system load: (a) undersubscribed ($n{=}1$), (b) matched ($n{=}2$), and (c) oversubscribed ($n{=}3$).}
    \vspace{-3.0ex}
\label{fig:main-eval}
\end{figure*}

\smalltt{MobileNetV2} is a video analytics pipeline in which each client process performs basic image preprocessing %—resizing, normalization, and color conversion—
before sending the data to convolutional neural network (CNN) inference service. The resulting labels are postprocessed for remote rendering. Each request corresponds to a batch of 80 preprocessed images, totaling 160MB. We use Intel AMX and `int8` quantized model.
\smalltt{XGBoost} is a predictive analytics pipeline where the client constructs feature vectors from structured tabular data and offloads inference to the server. Each request contains 200{,}000 rows from the breast cancer dataset, totaling approximately 25MB.
\smalltt{PageRank} is a graph analytics pipeline where each request contains a graph with 1 million vertices and 10 million edges, totaling 76MB. The computation exhibits high spatial and temporal locality due to repeated accesses to vertex rank values across iterations.
\smalltt{MilvusDB} is a similarity search pipeline where each request submits 200 image embeddings extracted via \smalltt{MobileNetV2} (1280-dimensional, float32, ~1MB) for approximate nearest neighbor search in a vector database. It was included to represent applications with large server-side responses, as each embedding yields top-3 similar images. \smalltt{Vision Transformer (ViT)} is a deep learning analytics pipeline for inference on image data. Each client request submits a batch of 200 preprocessed images, each with a resolution of 384x384 pixels. This model was included to represent modern, computationally intensive transformer-based workloads that require GPU acceleration.
%These benchmarks reflect real world deployment scenarios with differing pipeline structures, data volumes, and system bottlenecks.
%
% \smalltt{PageRank} is used to model a graph analytics workload, where a central server node performs iterative computation over large graphs and transmits the resulting rank vectors back to clients. This workload is characterized by large memory footprints and repeated memory accesses, placing sustained pressure on memory bandwidth. As such, it provides an opportunity to evaluate the effectiveness of DSA's buffer reuse and throughput-oriented design.
%
% To clarify the pipeline structure of each workload,
\autoref{tab:pipeline-detail} summarizes the major operations performed in each of the preprocessing, processing, and postprocessing stages. %This breakdown reflects realistic application scenarios and ensures that the evaluation is grounded in representative deployment conditions rather than synthetic benchmarks.

\noindent{\bf Performance Characteristics. } Our evaluation covers both CPU-only and GPU-accelerated settings.
In the CPU-only case, compute execution dominates total runtime,
and inter-process communication (IPC) accounts for only about 1\% of latency.
With GPU computation, faster compute phases make IPC and data movement a
larger portion of end-to-end latency, amplifying the benefits of memory
acceleration. This trend aligns with prior work~\cite{pocket, mlperf},
which reports 5-20$\times$ throughput gains in GPU-based execution where
data movement becomes a bottleneck.
Even in the CPU-only setup, where IPC is a small fraction, Rocket still
achieves substantial improvements through efficient IPC handling, lower
memory bus contention, reduced CPU cycles, and improved overlap between
compute and data transfer.

\subsection{Impact to Latency and Throughput}

\autoref{fig:main-eval} shows how DSA execution modes and cache injection affect end-to-end performance for the benchmarks.
We evaluate under three load conditions: (a) undersubscribed ($n{=}1$), (b) matched ($n{=}2$), and (c) oversubscribed ($n{=}3$). Each group reports execution time breakdown (top), throughput (middle), and latency improvement over the CPU-only baseline (bottom). The synchronous DSA configuration mirrors DTO in IPC stack and we additionally added DTO as a baseline for the \smalltt{MilvusDB} and the \smalltt{ViT} benchmarks.
Also, as the fraction of IPC communication increases within the end-to-end execution time, the performance benefits of offloaded IPC become more pronounced.
When running the Vision Transformer on GPUs rather than CPUs, we observe that the advantages of using DSA are significantly amplified.
This suggests that in multimodal or multi-compute-engine environments, the benefits of hardware-assisted IPC can become even more substantial.

\begin{figure}[h!]      % h: here, t: top, b: bottom, p: page of floats
  \centering
  \hspace{-0.5em}
  \includegraphics[
    width=1.00\linewidth,
    keepaspectratio
  ]{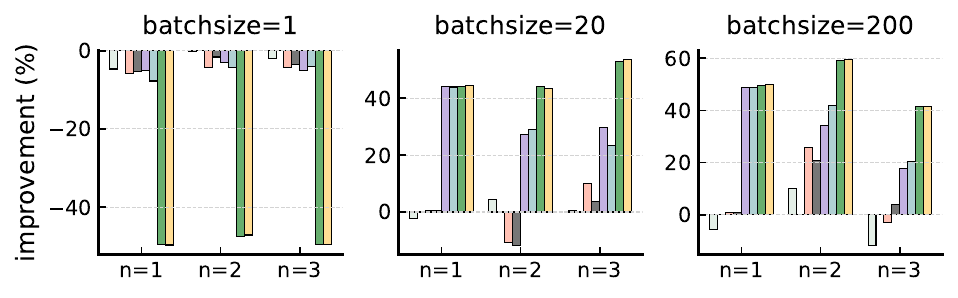}
  \includegraphics[width=1.05\linewidth]{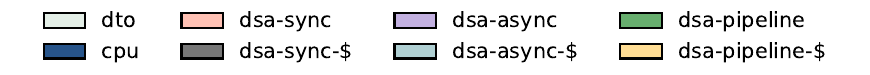}
\caption{
Latency improvement and the impact of data transmission size on optimal mode selection. 
(One input $\approx$ 600KB.) Even for the same application, the configuration parameters that yield the best performance can vary depending on the batch size, due to changes in the underlying data volume.}

% \ada{what are these data sizes not clear, is memcpy \% time diff for these? explicitly why were these experiments performed} \misun{how about now?}
    % \vspace{-4.5ex}
  \label{fig:batch-size}
\end{figure}

\begin{figure}[t]
  \centering
  \includegraphics[width=0.90\columnwidth]{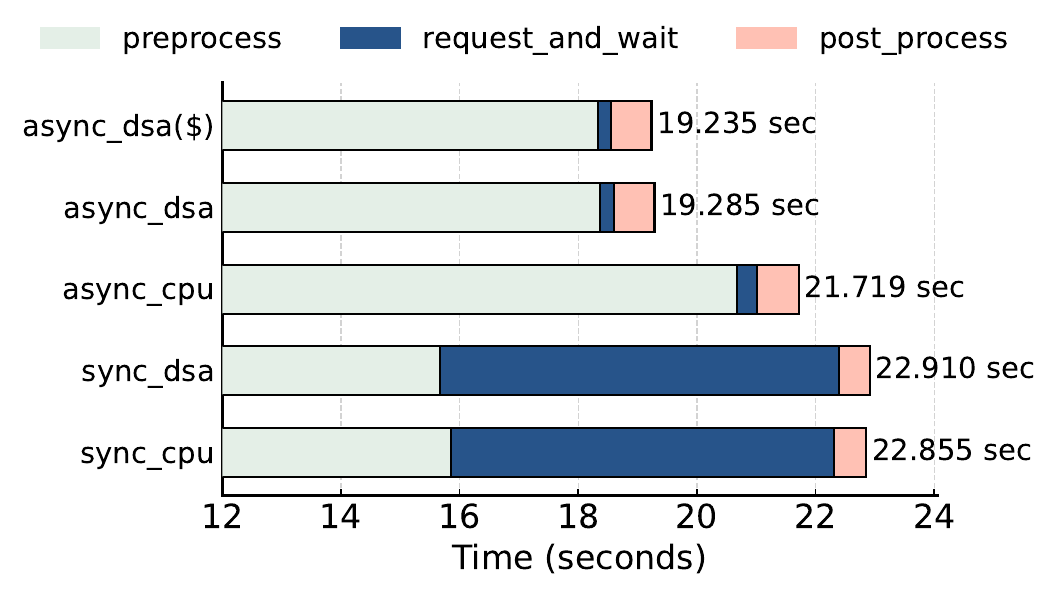}
  \vspace{-1.0ex}
  \caption{End-to-end latency decomposition across devices and execution modes (\texttt{MobileNetV2}).}
  %\ada{for which benchmark? do we know whether other benchmarks exhibit similar trend?}.}
  % \ada{can you add a line with sync-DSA? anyways need to make sure observation about gains are not just from switch to async is highlighted}
  \vspace{-5.5ex}
  \label{fig:async-cpu}
\end{figure}

\autoref{fig:async-cpu} shows that part of the latency reduction in asynchronous modes arises from changing the execution model itself, as seen in the improvement from \smalltt{sync\_cpu} to \smalltt{async\_cpu}. However, the majority of the benefit comes from DSA-based offloading, with \smalltt{async\_dsa} significantly outperforming \smalltt{async\_cpu}. This indicates that while asynchronous execution helps reduce idle time, the main contributor to performance gains is the reduced overhead and improved efficiency enabled by DSA.

% The model is executed using ONNX Runtime in a C++ environment, allowing for precise control over execution models and hardware interactions. We evaluate throughput, latency, and hardware events under different execution models, including synchronous, asynchronous, and pipelined DSA offloading. By capturing both high-level application performance and low-level system behavior, this evaluation provides insights into how DSA can be effectively leveraged in practical inference workloads, where large-scale memory movement is a critical bottleneck.

% \noindent
% \textbf{Benchmark Setup.}\hspace*{0.5em}
% When applying an asynchronous programming models, we observed throughput gains of up to 15\% and latency improvements of about 10\% relative to the baseline. Furthermore, we achieved over a 20\% reduction in CPU cycles and instruction counts, underscoring the potential of DSA-based IPC. The following subsection provides a more detailed analysis of the empirical observation about the source of these improvement in the hardware event level.

% \vspace{\baselineskip}

\noindent
\textbf{Comparison with Existing Software Support.} \hspace*{0.5em}
% \ada{make sure to update, now all have comparison with DTO}
In \autoref{fig:main-eval}, our evaluation includes eight configurations including \smalltt{dto} (vendor-supported). DTO operates synchronously with a single static configuration and offers no support for fine-grained tuning~\cite{dto}. For fairness, we used DTO with a reasonable configuration that enables hardware completion checks and balanced CPU-DSA work division (\texttt{AUTO\_ADJUST=1}, \texttt{WAIT\_METHOD=UMWAIT}). In our evaluation, DTO consistently underperforms in both throughput and latency, even falling behind the baseline without DSA. This stems from its indiscriminate use of DSA, forcing offloading even when it is counterproductive -- resulting in queuing delays and degraded performance where CPU-based copying would be more efficient. While DTO requires no code changes and is easy to adopt, it lacks the flexibility to apply DSA selectively based on workload demands.

\noindent
\textbf{Impact of Execution Modes.}\hspace*{0.5em}
\autoref{fig:main-eval} shows that asynchronous modes (\smalltt{async} and \smalltt{pipelined}) consistently outperform the synchronous baseline, confirming the design goal of reducing idle time through non-blocking execution.

\smalltt{Sync} mode, functionally similar to DTO, performs blocking offload and exhibits limited performance gains compared to CPU memcpy, highlighting the overheads of synchronous DSA usage.
\smalltt{Pipelined} mode batches requests and defers completion tracking, reducing polling and synchronization overhead. Since responses are rarely needed immediately, this deferred model improves efficiency.
Between the two asynchronous modes, \smalltt{async} offers lower per-request latency, while \smalltt{pipelined} yields higher throughput. This distinction is consistent across workloads. However, \smalltt{MobileNetV2} shows longer preprocessing delays than \smalltt{XGBoost} (\autoref{fig:main-eval}) in asynchronous modes due to higher memory contention, suggesting that CPU-DSA parallelism is more effective when CPU workloads are not memory-bound. Supporting all three modes allows \sys users to adapt their IPC implementation to their varying latency and throughput requirements.

\noindent
\textbf{Impact of Batch Size (Data Transfer Size).}\hspace{0.5em}
As shown in \autoref{fig:batch-size}, no single execution mode consistently outperforms others across all cases. When the data transfer size is small, the \smalltt{pipelined} mode performs the worst, but it becomes the most effective once the transfer size exceeds a certain threshold. We also observe that, for small transfers, a CPU-only pipeline without DSA yields the best performance.
Even in these cases, static DSA adoption via DTO often resulted in degraded performance.

% \vspace{\baselineskip}
\noindent
\textbf{Impact of Cache Injection.}\hspace*{0.5em}
% \ada{don't repeat again what/how some feature works, jump into what's new in terms of evaluation. remove the paragraph.}
% \autoref{sec:tradeoffs} highlighted that DSA's cache-bypassing behavior introduces cold-cache effects, increasing access latency when data is reused shortly after transfer. To mitigate this, \sys introduces an optional cache injection mechanism that routes data into the LLC during DSA transfers. This feature is selectively enabled based on workload context to avoid unnecessary pollution.
%Our evaluation confirms %\ada{
%the importance of providing explicit control over use of cache injection with IPC. %}
% that this design decision effectively balances latency and cache efficiency.
% As shown in \autoref{fig:main-eval}
The results show that cache injection improves latency in low-contention scenarios. For example, under single-threaded execution (\smalltt{MobileNetV2}), enabling injection in \smalltt{async} mode yields noticeable latency improvements, as the cache remains available for immediate reuse.
However, in multi-threaded configurations, with long reuse distance (likely from pipelined mode), or with data size of saturating LLC (likely to experience spatial/temporal cache contention), the same mechanism degrades performance. Simultaneous injection from multiple threads oversubscribes the cache, leading to eviction and lower hit rates. %In these cases, both latency and throughput drop below the baseline configuration with injection disabled.
These results validate that selective cache injection -- rather than naive always-on injection -- is a useful feature to tune IPC performance. % and \sys's adaptive strategy aligns with the reuse-aware design.
%  outlined in Section~\ref{sec:design-internals}, and avoids the pitfalls observed in prior work that lacked dynamic control.

% \vspace{\baselineskip}
\noindent
\textbf{Impact of Oversubscription of Cores.}\hspace*{0.5em}
%Throughout all experiments (\autoref{fig:main-eval}, \autoref{fig:main-eval-xgboost}) while the number of threads was varied, the number of CPU cores was fixed at two for the server and one per client instance to evaluate DSA's impact in both CPU-constrained and underutilized scenarios.
Across all system load scenarios -- undersubscribed, matched, and oversubscribed configurations -- asynchronous modes consistently outperforms the synchronous mode, with \smalltt{pipelined} optimizing for throughput and \smalltt{async} optimizing for latency.
Notably, as CPU resources became more constrained (i.e., with more threads), the relative latency improvement from DSA increases. This suggests that DSA's benefits are most pronounced in CPU-bound environments, where offloading memory operations frees up compute resources for application logic. In high-contention scenarios, reducing CPU-induced stalls and memory contention leads to greater performance gains than in lightly loaded environments.

\noindent
\textbf{Impact of Compute Accelerators.}\hspace*{0.5em}
We benchmarked the Vision Transformer (ViT) workload on both CPU and GPU configurations (\autoref{fig:main-eval}). \smalltt{ViT-cpu} exhibits high computational intensity. Furthermore, the total IPC time is very small ($\approx 0.2\%$). \sys provides the maximum benefits in the underloaded pipelined mode, when the end-to-end benefits are due to batched processing of requests, but otherwise it has negligible impact. Under higher loads ($n > 1$), there is stronger resource contention which overshadows the benefit of pipelining. On the CPU (second last row), compute-bound phases dominate the runtime. However, GPU provides over 10$\times$ increase in absolute end-to-end throughput, fundamentally shifting the performance bottlenecks. In GPU acceleration (bottom row), the compute bound phases shrink dramatically, causing IPC and data movement to account for a much larger fraction (about 10$\times$ higher than in the CPU configuration) of the end-to-end latency. This shift in performance bottlenecks towards IPC and data movement on the GPU is precisely why the pipelined mode helps improve overall system performance, even under higher loads ($n > 1$).  Consequently, asynchronous and pipelined DSA modes of \sys yield substantial throughput improvements of up to 40\%, which is a significant gain on a baseline that is already an order of magnitude faster than the CPU-only configuration. This shift amplifies the importance of efficient data handling.

\subsection{On the Source of Performance Improvement}

\noindent
\textbf{Reduced CPU and Bus Cycles.}\hspace*{0.5em}
\smalltt{memcpy}-induced instruction overhead and memory bus contention is a key performance bottlenecks in synchronous CPU and DSA execution. \sys addresses these issues by offloading transfers and overlapping them with computation, particularly in the \smalltt{async} and \smalltt{pipelined} modes.
\begin{figure}[h!]
  \centering
  \includegraphics[width=1.0\columnwidth]{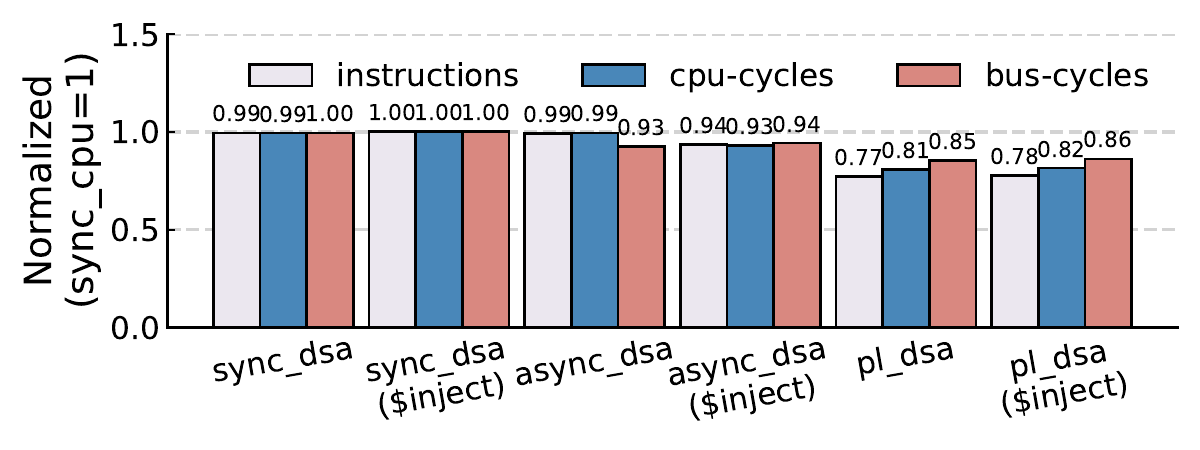}
  \vspace{-4.0ex}
  \caption{Normalized instruction counts, CPU cycles, and bus cycles with DSA-based offloading, relative to the synchronous CPU baseline.}
  % \vspace{-3.0ex}
  \label{fig:instr}
\end{figure}
% \ada{what's this figure, where is 23 coming from? and what happens in the sync DSA case?}
\autoref{fig:instr} shows that \sys reduces instruction count and CPU cycles by up to 23\% compared to the synchronous baseline. The largest gains appear in \smalltt{pipelined} mode, where burst submission and deferred synchronization minimize idle time and improve hardware utilization.
Reduced bus activity further indicates lower contention and more efficient bandwidth use.
These results %validate \sys’s concurrency-oriented design and
confirm
\sys's effectiveness in mitigating the CPU and bus bottlenecks discussed
in Section~\ref{sec:tradeoffs}. Compared to CPU \smalltt{memcpy} or blocking DSA like DTO, \sys's pipelined strategy offers higher efficiency for memory-bound workloads.

\begin{figure}[h]
  \centering
  \vspace{-1.5ex}
  \includegraphics[width=0.95\columnwidth]{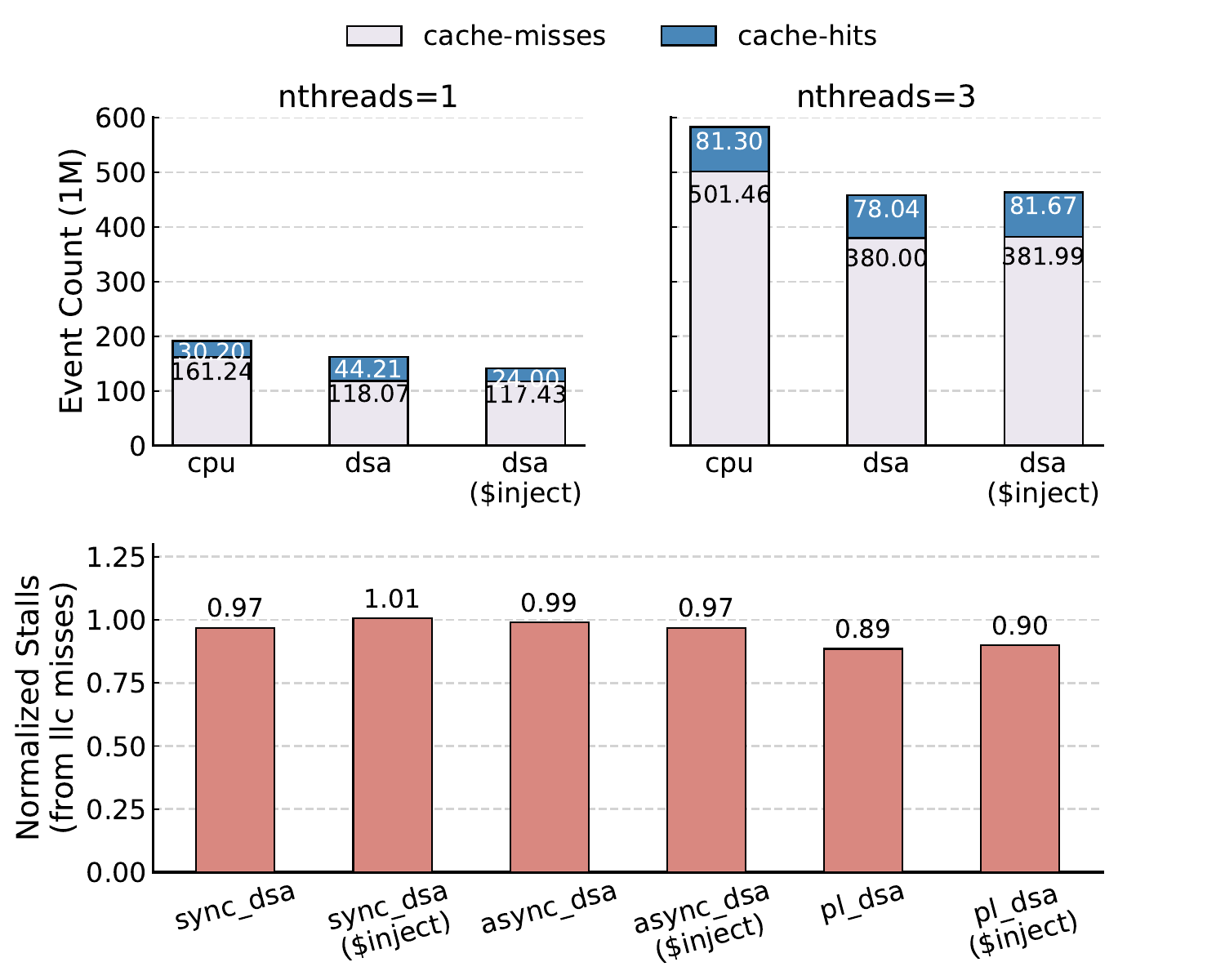}
  \caption{Cache hits, misses, and normalized CPU stalls with DSA-based offloading under matched load.
    % While DSA reduces overall cache activity, cache injection may introduce additional cache pollution, particularly in multi-threaded environments.
  }
    \vspace{-1.5ex}
  \label{fig:cache}
\end{figure}

\noindent \textbf{Impact on Cache Efficiency and CPU Stalls.}\hspace*{0.5em} Synchronization involving uncacheable memory accesses can reduce cache efficiency and introduce CPU stalls, especially in multi-threaded scenarios. \sys mitigates this by enabling cache injection only when reuse is likely, balancing latency and pollution.
\autoref{fig:cache} shows that DSA offloading reduces both cache hits and misses under matched-thread conditions, indicating lower overall cache activity compared to CPU-based \smalltt{memcpy}. This aligns with prior findings~\cite{cache-injection}, where cache injection lowers both metrics by reducing access volume via preemptive placement.
However, under higher thread counts, enabling cache injection increases cache accesses -- in the 3-thread case, due to poor reuse and interference. This leads to elevated reference churn, negating performance gains.
The lowest stall rate (0.89) occurs in \smalltt{pipelined} mode with two threads and no injection, supporting \sys’s strategy to disable injection under concurrent, deferred-access workloads.

\begin{figure}[h]
  \centering
  % \hspace{-2.5ex}
  \includegraphics[width=0.85\columnwidth]{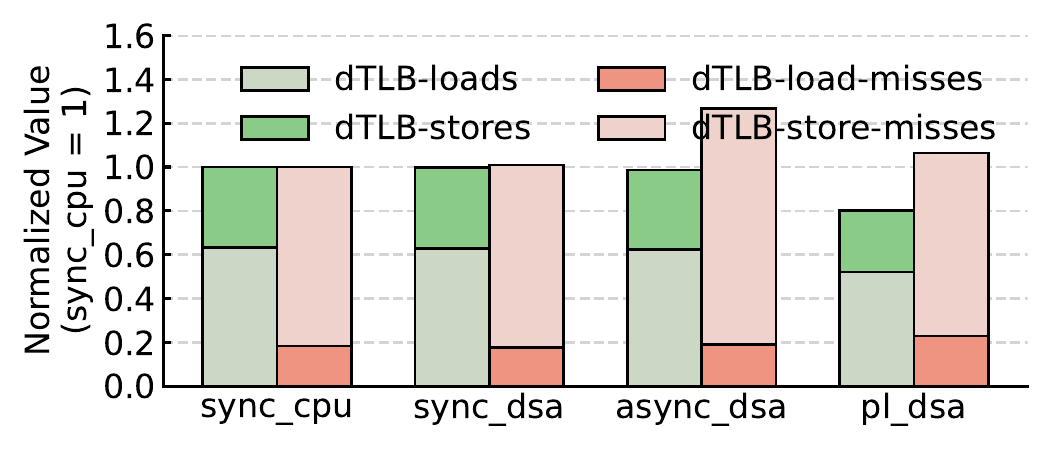}
  \caption{Normalized dTLB activity under matched load across CPU and DSA modes.}
  % \vspace{-4.0ex}
  \label{fig:dtlb}
\end{figure}

\noindent
\textbf{Impact on Memory Bandwidth and Bus Contention.}\hspace*{0.5em}
Offloading can disrupt the memory hierarchy and reduce TLB efficiency, particularly in asynchronous execution.
As shown in \autoref{fig:dtlb}, dTLB miss rates increase under \smalltt{async} and \smalltt{pipelined} modes, since address translation is handled by the IOMMU, bypassing CPU-resident TLBs. This indicates that while offloading reduces CPU load, it limits reuse of translation entries and weakens locality.
\sys's \smalltt{pipelined} mode mitigates these effects by batching memory operations, improving access predictability.
These results validate \sys's design trade-off -- sacrificing some TLB locality for parallelism.
% —
% and show that pipelined execution strikes a favorable balance for high-volume workloads.

\sys improves performance by addressing the system-level inefficiencies identified in \Cref{sec:tradeoffs}. Asynchronous modes eliminate CPU blocking, with \smalltt{pipelined} mode yielding up to 15\% higher throughput and 23 fewer instructions than the synchronous baseline. Cache injection enhances latency in single-threaded cases but is disabled under multithreading to prevent cache pollution. Batched pipelined transfers further alleviate memory subsystem pressure by reducing stall rates and bus contention versus naïve offloading.

These results show that \sys avoids common limitations in prior approaches, such as blocking execution or indiscriminate cache injection, by exposing an adaptive interface to adjusts the IPC behavior to workload characteristics. As a result, \sys makes more effective use of DSA than existing systems (e.g., DTO, idxd), while maintaining stable performance under concurrency, varying locality, and system-level contention.

\vspace{-2.0ex}
\section{Related Work}\label{sec:related}

\noindent
\textbf{DSA-Based Data Movement Acceleration.}\hspace*{0.5em}
Recent work has explored using Intel’s Data Streaming Accelerator (DSA) to offload memory-intensive operations across diverse domains. Prior efforts include benchmarking and characterization of DSA's microarchitectural behavior~\cite{asplos-kuper}, transparent application-level offloading via DTO~\cite{dto}, and practical use cases such as memory deduplication~\cite{para-ksm} and tiered memory systems~\cite{dsa-2lm}. These systems leverage DSA’s hardware features (e.g., descriptor batching and work queues) to improve efficiency in specific scenarios. While DTO simplifies usage by intercepting \texttt{memcpy} calls, it lacks programmability and fine-grained control. Our work builds on these foundations by using DSA for IPC and investigating how DSA’s performance characteristics change depending on memory conditions. Unlike prior work, \sys supports multiple execution modes and offers configurable trade-offs between latency and offload granularity, enabling better adaptation across workloads consisting of multiple processes.

\vspace{0.5em}
\noindent
\textbf{Data Movement Engines and Systems Support.}\hspace*{0.5em}
Prior work has explored accelerating data movement using traditional DMA engines (e.g., Intel I/OAT~\cite{ioat-ipdps}), RDMA-capable NICs, and specialized hardware like DMX~\cite{dmx} and MC$^2$~\cite{mc2}. These systems exploit asynchronous offloading to reduce CPU involvement and improve throughput across storage or accelerator pipelines. At the OS level, recent work such as \textit{Copier}~\cite{copier} elevates asynchronous copy operations into a coordinated kernel service, complementing hardware offload by managing copy-compute overlap system-wide. In contrast, Rocket focuses on intra-node IPC where the orchestration of memory movement must account for process boundaries, shared memory semantics, and responsiveness. Our design exposes fine-grained controls to match workload-specific needs in IPC contexts. Nevertheless, several of \sys's coordination mechanisms -- such as mode switching, queueing, and cost-aware path selection -- could generalize to other emerging engines, enabling software runtime systems to better exploit their async capabilities.

\section{Limitations and Future Work}\label{sec:future}
\sys targets intra-node IPC and does not support inter-node communication. While this limits its direct deployment in distributed environments, its key design principles, such as async memory orchestration and cache-aware transfers, remain applicable to other hardware with memory offloading and cache/sync control. As systems adopt multi-node GPUs, pooled memory, and CXL-shared address spaces, the problems \sys addresses -- overlapping data movement and managing cache effects -- are increasingly relevant. However, emprically exploring \sys's applicability in inter-node runtimes is left to future work.

\sys is not self-tuning, though it offers reasonable defaults. It exposes critical parameters (e.g., transfer thresholds, polling budgets, cache policies) via manual interfaces. While our empirical characterization informs expert use, \sys lacks a runtime decision engine. We argue that identifying such parameters and understanding their workload interactions is a prerequisite to effective autotuning—this is the core contribution of \sys. Building tuners (e.g., rule-based, statistical, or ML-based) is orthogonal and remains future work; we are actively exploring this direction.

\section{Conclusion}
\label{sec:conclusion}
Hardware-assisted memory offloading is becoming a cornerstone of modern system design, decoupling data movement from CPU execution and enabling new forms of parallelism.
While such mechanisms promise substantial efficiency gains, realizing their potential requires software runtimes that can manage synchronization, visibility, and cache behavior coherently across process boundaries.
This paper presents \sys, a runtime that integrates Intel’s DSA into user-space IPC to demonstrate how hardware offload can be systematically exploited in software.
\sys introduces asynchronous execution modes, reuse-aware cache injection, and lightweight synchronization, turning low-level offload interfaces into configurable IPC primitives.
% \ada{check these against the actual results}
Evaluations across representative workloads show that \sys reduces instruction count by up to 22\%, improves throughput by up to 2.1$\times$, and latency by 72\%, demonstrating that hardware offload can directly translate into end-to-end gains when tightly coupled with runtime control.
By bridging hardware acceleration and software orchestration within IPC, \sys shows a practical path toward treating memory offloading as a first-class system capability.

\bibliographystyle{IEEEtranS}
\bibliography{paper}
%%%%%%%%%%%%%%%%%%%%%%%%%%%%%%%%%%%%

\end{document}